\documentclass[12pt]{article}
\usepackage{amsmath,amssymb,latexsym,bbold,epsfig,axodraw}
\usepackage{amsfonts}
\usepackage{bbm}
\usepackage{dsfont}
\usepackage{slashed}
\usepackage{cite}
\usepackage[usenames,dvipsnames]{xcolor}
\definecolor{lightgray}{gray}{0.85}

\renewcommand{\vec}[1]{{\mathbf{#1}}}

\usepackage[colorlinks=true,linkcolor=red,citecolor=green,urlcolor=cyan,
bookmarks=true,bookmarksopen=true,pdfpagemode=None,pdfstartview=FitH]{hyperref}

\newcommand{\be}{\begin{equation}}
\newcommand{\ee}{\end{equation}}
\newcommand{\bea}{\begin{eqnarray}}
\newcommand{\eea}{\end{eqnarray}}
\begin{document}
\title{
\vglue -0.3cm
\vskip 0.5cm
\Large \bf
Majorana neutrinos and other Majorana particles:\\ 
Theory and experiment}
\author{
{Evgeny~Kh.~Akhmedov%
\footnote{Also at the National Research Centre
Kurchatov Institute, Moscow, Russia}
\thanks{email: \tt 
akhmedov@mpi-hd.mpg.de}
} \\
{\normalsize\em Max-Planck-Institut f\"ur Kernphysik,
Saupfercheckweg 1} \\ {\normalsize\em D-69117 Heidelberg, Germany
\vspace*{0.15cm}}
} 
\date{}
\maketitle
\thispagestyle{empty}
\vspace{-0.8cm}
\begin{abstract}
This is a somewhat modified version of Chapter 15 of the book {\em``The 
Physics of Ettore Majorana''}, by Salvatore Esposito with contributions 
by Evgeny Akhmedov (Ch.~15) and Frank Wilczek (Ch.~14), Cambridge University 
Press, 2014. 
\end{abstract}

\vspace{-0.8cm}
\maketitle
\thispagestyle{empty}

\vspace{1.cm}

\newpage

What are Majorana particles? These are massive fermions that are their 
own antiparticles. In this chapter we will concentrate on spin-1/2 
Majorana particles, though fermions of higher spin can also be of Majorana 
nature. Obviously, Majorana particles must be genuinely neutral, i.e.\ 
they cannot possess any conserved charge-like quantum number that would 
allow one to discriminate between the particle and its antiparticle. In 
particular, they must be electrically neutral. Among the known spin-1/2 
particles, only neutrinos can be of Majorana nature. Another known 
quasi-stable neutral 
fermion, the neutron, has non-zero magnetic moment which disqualifies it 
for being a Majorana particle: the antineutron exists, and its magnetic 
moment is negative of that of the neutron%
\footnote{On could have also argued that neutron and antineutron are 
distinguished by their baryon number (+1 and $-1$, respectively), but 
conservation of baryon number is not an exact symmetry of Nature.}.

Neutrinos are exactly massless in the original version of the standard model 
of electroweak interaction, and are massive Majorana particles in most its 
extensions. Although massive Dirac neutrinos is also a possibility, most 
economical and natural models of neutrino mass lead to Majorana neutrinos. 
Since only massive neutrinos can oscillate, the interest to the possibility 
of neutrinos being Majorana particles rose significantly after the first hints 
of neutrino oscillations obtained in the solar and atmospheric neutrino 
experiments. It has greatly increased after the oscillations were 
firmly established in the experiments with solar, atmospheric, accelerator and 
reactor neutrinos \cite{rev1,rev2,StruViss}. 
In addition to being the simplest and most economical possibility, Majorana 
neutrinos bring in two important added bonuses: they can explain the 
smallness of the neutrino mass in a very natural way through the so-called 
seesaw mechanism, and can account for the observed baryon asymmetry of the 
Universe through `baryogenesis via leptogenesis'. We shall discuss both in 
this chapter. 

In the limit of vanishingly small mass the difference between Dirac 
and Majorana fermions disappears.
Therefore the observed smallness of 
the neutrino mass makes it very difficult to discriminate 
between different types of massive neutrinos, and it is not currently known 
if neutrinos are Majorana or Dirac particles. 
The most promising means of finding this out is through the experiments 
on neutrinoless double beta decay. Such experiments are currently being 
conducted in a number of laboratories. 

In this chapter we review the properties of Majorana neutrinos and other 
Majorana particles. We start with discussing Weyl, Dirac and Majorana 
fermions and comparing the Dirac and Majorana mass terms. We then proceed to 
discuss C, P, CP and CPT properties of Majorana particles in sec.\ 15.2.  
This is followed by a discussion of mixing and oscillations of 
neutrinos in the Majorana and general Dirac + Majorana cases in sec.\ 15.3. 
In sec.\ 15.4 we discuss the seesaw mechanism of the neutrino mass generation, 
which is the leading candidate for the explanation of the 
smallness of the neutrino mass. Next, we consider electromagnetic properties 
of Majorana neutrinos in sec.\ 15.5. Section 15.6 contains a brief discussion 
of Majorana particles predicted by supersymmetric theories. In sec.\ 15.7 
we review theoretical foundations and the experimental status of the 
neutrinoless $2\beta$-decay as well as of other processes that could 
distinguish between Majorana and Dirac neutrinos. 
Our next topic is baryogenesis via leptogenesis due to lepton number 
violating processes caused by Majorana neutrinos (sec.\ 15.8). 
Finally, in sec.\ 15.9 we collect a few assorted remarks on Majorana particles
and in sec.\ 15.10 summarize the main points of our discussion.

\subsection*{\label{sec:th}15.1\, Weyl, Dirac and Majorana fermions} 

\noindent
Being dissatisfied with the interpretation of antifermions as holes in the 
Dirac sea, in his famous paper \cite{Maj} Majorana sought to cast the Dirac 
equation in a form that would be completely symmetric with respect to 
particles and antiparticles. 
He succeeded to do that by finding a new form of the Dirac equation, in which 
all coefficients were real. While it  
led to only formal improvement 
for charged fermions, 
the Majorana form of the Dirac equation opened up a very important 
new possibility for neutral ones -- they can be their own antiparticles. The 
Majorana particles are thus fermionic analogues of genuinely neutral bosons, 
such as the $\pi^0$-meson or the photon.   

Recall that a free spin-1/2 fermion field in general satisfies the Dirac 
equation%
\footnote{We use the natural units $\hbar=c=1$ and assume 
summation over repeated indices in this chapter.}
\be
(i\gamma^\mu \partial_\mu - m)\psi(x)=0\,,
\label{eq:D1}
\ee
where $\partial_\mu\equiv \partial/\partial x^\mu$, $\psi(x)$ is a 4-component 
spinor field, $m$ is the mass of the fermion, 
and $\gamma^\mu$ ($\mu=0, 1, 2, 3)$ are $4\times 4$ matrices 
satisfying  
\be
\{\gamma^\mu, \gamma^\nu\}=2g^{\mu\nu}\cdot \mathbbm{1}\,,\qquad
\gamma^0 \gamma^{\mu\dag} \gamma^0=\gamma^\mu\,,
\label{eq:D2}
\ee
with $g^{\mu\nu}=diag(1, -1,-1,-1)$ and $\mathbbm{1}$ being the flat 
space-time metric tensor and the $4\times 4$ unit matrix, respectively. 
Note that the Dirac equation 
(\ref{eq:D1}) can be cast in the Schr\"odinger form $i(\partial/\partial t)
\psi(x)=H_D\psi(x)$, where $H_D=-i\gamma^0\boldsymbol{\gamma}\cdot
\boldsymbol{\nabla}+\gamma^0 m$. The first equality in eq.~(\ref{eq:D2}) 
follows from the requirement that the solutions of the Dirac equation obey  
the usual dispersion law of free relativistic particles $E^2=\vec{p}^2+m^2$, 
while the second equality follows from hermiticity of the Dirac Hamiltonian 
$H_D$. In addition to the matrices $\gamma^\mu$, a very important role is 
played by the matrix $\gamma_5\equiv i\gamma^0 \gamma^1 \gamma^2 \gamma^3$ 
which satisfies 
\be
\{\gamma_5, \gamma^\mu\}=0\,,\qquad \gamma_5^\dag=\gamma_5\,, \qquad
\gamma_5^2=\mathbbm{1}\,.
\label{eq:D3}
\ee
There are infinitely many unitarily equivalent representations of the Dirac  
matrices. In this chapter, unless otherwise specified, we will use the 
so-called chiral (or Weyl) representation 
\be
\gamma^0=\left(\begin{array}{cc} {0} & \mathds{1}\\\mathds{1} & {0}
\end{array}\right)\,,\qquad 
\gamma^i=\left(\begin{array}{cc} {0} & 
\sigma^i\\ -\sigma^i & {0}
\end{array}\right)\,,\qquad
\gamma_5=\left(\begin{array}{cc} -\mathds{1} & 0 \\ 
0 & \mathds{1} 
\end{array}\right)\,,
\label{eq:D4}
\ee
where $\mathds{1}$ and $0$ are the unit and zero $2\times 2$ matrices, and 
$\sigma^i$ ($i=1, 2, 3$) are the standard Pauli matrices. 

The left-handed and right-handed chirality projector operators $P_{L,R}$ 
are defined as 
\be
P_L = \frac{\mathbbm{1}-\gamma_5}{2}\,, \quad\quad
P_R = \frac{\mathbbm{1}+\gamma_5}{2}\,.
\label{eq:PLPR}
\ee
They have the following properties:
\be
P_L^2=P_L\,,\quad P_R^2=P_R\,, \quad P_L P_R=P_R P_L=0\,, \quad P_L+P_R=
\mathbbm{1}\,.
\label{eq:rel1}
\ee
Any spin-1/2 fermion field $\psi$ can be decomposed into the sum of its 
left-handed and right-handed components according to 
\be
\psi=\psi_L + \psi_R\,,\quad\mbox{where}\quad \psi_{L,R}=P_{L,R}\psi=
\frac{\mathbbm{1}\mp \gamma_5}{2}\psi\,.
\label{eq:decomp}
\ee
Note that the chiral fields $\psi_{L,R}$ are eigenstates of 
$\gamma_5$: $\gamma_5 \psi_{L,R}=\mp \psi_{L,R}$. 
The terms `left-handed' and `right-handed' originate from the fact that 
for relativistic particles chirality almost coincides with helicity defined as 
the projection of the spin of the particle on its momentum. More precisely, in 
the relativistic limit, for positive-energy solutions of the Dirac equation 
the left- and right-handed chirality fields approximately coincide with those 
of negative and positive helicity, respectively. The helicity projection 
operators are
\be
P_{\pm} = \frac{1}{2}\left(1 \pm\frac{\mbox{\boldmath $\sigma$}{\bf p}}
{|\bf p|}\right). 
\label{eq:helic}
\ee
They satisfy relations similar to (\ref{eq:rel1}).  
For a free fermion, helicity is conserved but chirality in general is not; 
it is only conserved in the limit $m=0$, when it coincides with helicity.
However, for relativistic particles chirality is nearly conserved, and 
the description in terms of chiral states is useful.

For our discussion we will need the particle - antiparticle conjugation 
operator $\hat{C}$. Its action on a fermion field $\psi$ is defined as  
\be
\hat{C}: \; \psi \to \psi^c = {\cal C} \bar{\psi}^T\,, 
\label{eq:C}
\ee
where $\bar{\psi}\equiv\psi^\dag \gamma^0$ is the adjoint field and 
the matrix ${\cal C}$ satisfies 
\be
{\cal C}^{-1}\gamma^\mu {\cal C}=-\gamma^{\mu T}\,,\qquad 
{\cal C}^{-1}\gamma_5 {\cal C}=\gamma_5^T\,,\quad\qquad 
{\cal C}^\dagger={\cal C}^{-1}=-{\cal C}^*\,. 
\label{eq:C1}
\ee
Note that the second equality here follows from the first one and the 
definition of $\gamma_5$.  
For free particles, the $\hat{C}$-conjugate field $\psi^c(x)$ satisfies the 
same Dirac equation as $\psi(x)$. 
Some useful relations that follow from (\ref{eq:C}) and (\ref{eq:C1}) are 
\be
(\psi^c)^c=\psi\,,\qquad 
\overline{\psi^c}=-\psi^T {\cal C}^{-1}\,,\qquad
\overline{\psi_k} \psi_i^c=\overline{\psi_i} \psi_k^c\,,\qquad
\overline{\psi_k} A \psi_i =\overline{\psi_i^c} ({\cal C} A^T {\cal C}^{-1})
\psi_k^c\,,
\label{eq:C2}
\ee
where $\psi$, $\psi_i$, $\psi_k$ are anticommuting 4-component fermion fields
and $A$ is an arbitrary $4\times 4$ matrix. 
Note that the third equality in (\ref{eq:C1}) means that the matrix ${\cal C}$ 
is antisymmetric. 
In the representation 
(\ref{eq:D4}) as well as in a number of other representations of the Dirac 
matrices one can choose e.g. ${\cal C}=i\gamma^2\gamma^0$. In this case 
${\cal C}$ is real, ${\cal C}^{-1}=-{\cal C}$, and $\overline{\psi^c}=\psi^T 
{\cal C}$. For future use, we give here the expressions 
${\cal C} A^T {\cal C}^{-1}$ for several matrices $A$:
\be
{\cal C} \gamma^{\mu T}{\cal C}^{-1}=-\gamma^\mu\,,\qquad\quad~~~\;
{\cal C} (\gamma^\mu\gamma_5)^T {\cal C}^{-1}=
\gamma^\mu\gamma_5\,,
\quad\;\;\,
\nonumber \\
\ee
\be
{\cal C}(\sigma^{\mu\nu})^T{\cal C}^{-1}=-{}\sigma^{\mu\nu}\,,\qquad~\;
{\cal C}(\sigma^{\mu\nu}\gamma_5)^T {\cal C}^{-1}=-{}\sigma^{\mu\nu}\gamma_5\,,
\;
\label{eq:C2a}
\ee
where $\sigma^{\mu\nu}\equiv \frac{i}{2}[\gamma^\mu,\gamma^\nu]$.

Using the anticommutation properties of the Dirac $\gamma$-matrices it is 
easy to see that, acting on a chiral field, $\hat{C}$ flips its chirality: 
\be
\hat{C}: \; \psi_L\to(\psi_L)^c=(\psi^c)_R\,, \quad\quad
\psi_R\to (\psi_R)^c=(\psi^c)_L\,,
\label{eq:C3}
\ee
i.e.\ the antiparticle of a left-handed fermion is right-handed. This fact 
plays a very important role in the theory of Majorana particles.

The particle - antiparticle conjugation operation $\hat{C}$ must not be 
confused with the charge conjugation operation C which, by
definition, flips all the charge-like quantum numbers of a field (electric 
charge, baryon number $B$, lepton number $L$, etc.) but leaves all the other 
quantum numbers (including chirality) intact. In particular, charge conjugation 
would take a left-handed neutrino into a left-handed antineutrino that does 
not exist, which is a consequence of maximal C-violation in weak 
interactions. At the same time, $\hat{C}$-conjugation converts a left-handed 
neutrino into a right-handed antineutrino which does exist and is the 
antiparticle of the left-handed neutrino. 

\vspace*{1.5mm}
\colorbox{lightgray}
{\parbox[t]{15cm}{\footnotesize 
A little caveat should be added to the above. 
Strictly speaking, a 
particle and its antiparticle are related by the CPT transformation, as only 
this combination of 
the charge conjugation C, space parity P and time reversal T 
is exactly conserved in any 
`normal' theory (i.e.\ local Poincar\'e invariant Lagrangian quantum field 
theory with the usual relation between spin and statistics). However, the 
CP conjugation does essentially the same job as far as (typically very small) 
effects of CP-violation can be neglected. The $\hat{C}$ conjugation introduced 
in eq.~(\ref{eq:C}) acts very similarly to the CP conjugation as it flips all 
the non-zero charges of the fermion as well as its chirality, which is odd 
under P transformation. We discuss these points in more detail in sec.\ 15.2. 
It should be added that when we say that the charge conjugation C flips the 
baryon and lepton numbers of the particles we assume that these numbers are 
well defined, i.e.\ that small effects of 
$B$ and $L$ violation can be ignored. 
}}

\vspace*{1.8mm}

Let us now return to the discussion of the Dirac equation. Adopting the 
Weyl representation of the Dirac $\gamma$-matrices (\ref{eq:D4}) and 
writing the 4-component spinor field $\psi(x)$ in terms of the 2-component 
spinors $\phi(x)$ and $\xi(x)$ as 
\be
\psi=\left(\begin{array}{c}\phi \\ \xi \end{array}\right),
\label{eq:D5}
\ee
one can rewrite the Dirac equation (\ref{eq:D1}) as a set of two coupled 
equations for $\phi$ and $\xi$: 
\bea
(i\partial_0-i\boldsymbol{\sigma}\cdot\boldsymbol{\nabla})\phi-m\xi=0\,,
\nonumber \\
(i\partial_0+i\boldsymbol{\sigma}\cdot\boldsymbol{\nabla})\xi-m\phi=0\,.
\label{eq:D6}
\eea
{}From the expression for $\gamma_5$ in eq.~(\ref{eq:D4}) and eq.~(\ref{eq:D5}) 
one obtains
\be
\psi_L=\left(\begin{array}{c} \phi \\ 0 \end{array}\right),\qquad
\psi_R=\left(\begin{array}{c} 0 \\ \xi \end{array}\right),
\label{eq:chiral}
\ee  
i.e.\ the 2-component spinor fields $\phi$ and $\xi$ determine, respectively, 
the left- and right-handed components of the 4-component field $\psi$. 
Thus, the chiral fields are actually 2-component rather than 4-component 
objects. 

{}From eq.~(\ref{eq:D6}) it follows that in the limit $m=0$ the equations for 
$\phi$ and $\xi$ decouple, i.e. the left-handed and right-handed components 
of $\psi$ evolve independently. The resulting equations are called the Weyl 
equations, and the corresponding chiral solutions describe massless spin-1/2 
particles called Weyl fermions. At the same time, as follows 
from~(\ref{eq:D6}),
to describe a massive fermion one needs both left-handed and right-handed 
chiral fields. 

The latter statement can also be demonstrated as follows. The Dirac equation 
for a free spin-1/2 particle can be obtained as the Euler-Lagrange equation 
applied to the Dirac Lagrangian
\be
{\cal L}=\bar{\psi}(i\gamma^\mu \partial_\mu - m)\psi\,.  
\label{eq:L1}
\ee
The mass term of this Lagrangian can be written as 
\be
-{\cal L}_m=m\bar{\psi}\psi=m\overline{(\psi_L+\psi_R)}(\psi_L+\psi_R)
=m (\overline{\psi_L}\psi_R+\overline{\psi_R}\psi_L)
\,,
\label{eq:mass}
\ee
i.e.\ only the cross terms survive while the $\overline{\psi_L}\psi_L$ 
and $\overline{\psi_R}\psi_R$ terms vanish identically. Thus, one needs both 
left-handed and right-handed chiral fields to construct the mass term of the 
Lagrangian, and a massive fermion field must be a sum of them: 
$\psi=\psi_L+\psi_R$.%
\footnote{Note that the kinetic term of the Lagrangian (\ref{eq:L1}) is 
decomposed as $\bar{\psi} i\gamma^\mu \partial_\mu\psi=
\bar{\psi}_L i\gamma^\mu \partial_\mu\psi_L+\bar{\psi}_R i\gamma^\mu 
\partial_\mu\psi_R$. In other words, 
for each chiral component the kinetic term can be written 
separately and therefore it does not require the existence of both 
components.
}

Now, there are essentially two possibilities. First, the right-handed
component of a massive field can be completely independent of the left-handed 
one; in this case we have a Dirac field. The second, and the most important 
for us possibility, is based on the discussed above fact that the 
particle-antiparticle 
conjugate of a left-handed field is right-handed. Therefore the 
right-handed component of a massive spin-1/2 field can be 
just the $\hat{C}$ - conjugate of its left-handed component: $\psi_R=
(\psi_L)^c=(\psi^c)_R$, or 
\be
\psi=\psi_L+(\psi_L)^c=\psi_L+(\psi^c)_R\,.
\label{eq:psi1}
\ee
In this case we have a Majorana field; one can construct it with just one 
chiral field. From (\ref{eq:psi1}) it immediately follows that
the $\hat{C}$ - conjugate field coincides with the original one:
\be
\psi^c=\psi\,.
\label{eq:C4}
\ee
This means that particles associated with Majorana fields 
are genuinely neutral, i.e.\ they 
are their own antiparticles. 
The condition in eq.~(\ref{eq:C4}) is called the Majorana condition.

\vspace*{1.5mm}
\colorbox{lightgray}
{\parbox[t]{15cm}{\footnotesize 
In his paper \cite{Maj} Majorana found a representation of the 
$\gamma$-matrices in which they were all pure imaginary, so that the Dirac 
equation (\ref{eq:D1}) did not contain any complex coefficients. As a result, 
the equation admitted real solutions 
\be
\psi^*=\psi\,,
\label{eq:Majcond1}
\ee
which describe genuinely neutral particles. Eq.~(\ref{eq:C4}) 
generalizes the Majorana condition (\ref{eq:Majcond1}) to the case of an 
arbitrary representation of the $\gamma$-matrices (see e.g. \cite{Pal} for a 
formal proof). 

It is easy to see that the general self-conjugacy condition 
(\ref{eq:C4}) indeed reduces to (\ref{eq:Majcond1}) in the Majorana basis. 
In the Majorana representation the $\gamma$-matrices 
satisfying eq.~(\ref{eq:D2}) can be chosen as 
\be
\gamma_{\rm M}^0=\left(\begin{array}{cc}
0 & \sigma^2 \\
\sigma^2 & 0
\end{array}\right)\!,~~~~
\gamma_{\rm M}^1=i\left(\begin{array}{cc}
\sigma^3 & 0 \\
0 &\sigma^3
\end{array}\right)\!,~~~~
\gamma_{\rm M}^2=\left(\begin{array}{cc}
0 & -\sigma^2 \\
\sigma^2 & 0
\end{array}\right)\!,~~~~
\gamma_{\rm M}^3=-i\left(\begin{array}{cc}
\sigma^1 & 0 \\
0 & \sigma^1
\end{array}\right)\!,\nonumber
\ee
where the subscript M stands for the Majorana basis. All the $\gamma$-matrices 
are pure imaginary, as required (note that this representation 
is not unique). The matrix $\gamma_{5\rm M}$ is then 
\be
\gamma_{5 \rm M}^0=\left(\begin{array}{cc}
\sigma^2 & 0\\
0 & -\sigma^2 
\end{array}\right)\!.\nonumber
\ee
Notice that 
$\gamma_{\rm M}^0$ is antisymmetric, whereas $\gamma_{\rm M}^i$ ($i=1, 2, 3$) 
are symmetric; the particle-antiparticle conjugation matrix ${\cal C}$ 
satisfying eq.~(\ref{eq:C1}) can therefore be chosen as 
\be
{\cal C}_{\rm M}=-\gamma^0_{\rm M}=-\left(\begin{array}{cc}
0 & \sigma^2 \\
\sigma^2 & 0
\end{array}\right)\!.
\label{eq:CM}
\ee
{}From eq.~(\ref{eq:C}) we then find 
\be
\psi_{\rm M}^c={\cal C}_{\rm M}\bar{\psi}_{\rm M}^T=
{\cal C}_{\rm M}\gamma_{\rm M}^{0 T} \psi_{\rm M}^*=
-\gamma^0_{\rm M}\gamma_{\rm M}^{0 T} \psi_{\rm M}^*=
\psi_{\rm M}^*\,,
\label{eq:MajConj}
\ee
i.e.\ the condition that the particle is its own antiparticle $\psi^c=\psi$
reduces in the Majorana basis to 
the requirement that the field $\psi_{\rm M}$ be real. 
}}

\vspace*{2.8mm}

As was discussed above, to construct a massive Dirac field one needs two 
independent 2-component chiral fields, $\psi_L$ and $\psi_R$; 
this gives four degrees of freedom. In
contrast with this, a Majorana fermion has only two degrees of freedom, 
because its right-handed component is constructed from the 
left-handed one.
Thus, Majorana fermions are actually simpler and more economical constructions 
than the Dirac ones. 

While Majorana fields are essentially 2-component objects, it is often useful 
to write them in the 4-component notation, especially when considering 
processes in which Majorana particles participate along with Dirac ones.  
It is easy to see that in the chiral representation of the Dirac matrices 
the Majorana field can be written in the 4-component form as 
\be
\psi=\left(\!\!
\begin{array}{c}
\phi \\ -i\sigma^2\phi^*
\end{array}\!\!\right).
\label{eq:M}
\ee 
Indeed, from ${\cal C}=i\gamma^2\gamma^0$ and eq.~(\ref{eq:C}) we have 
\be
\psi^c=i\gamma^2\psi^*=\left(
\begin{array}{cc}
0 & i\sigma^2 \\
-i\sigma^2 & 0
\end{array}\right)
\left(\!\!
\begin{array}{c}
\phi^* \\ -i\sigma^2\phi
\end{array}\!\!\right)
=\left(\!\!
\begin{array}{c}
\phi \\ -i\sigma^2\phi^*
\end{array}\!\!\right)=\psi\,.
\label{eq:M1}
\ee

To understand better the difference between the Dirac and Majorana particles 
it is instructive to look at the expansions of their quantum fields  
in terms of the plane-wave modes.
Recall that for a Dirac field the expansion has the form
\be
\psi(x)=\int\frac{d^3p}{(2\pi)^3\sqrt{2E_{\vec{p}}}}\sum_s \left[
b_{s}(\vec{p}) u_{s}(\vec{p}) e^{-i p x} + d^\dag_{s}
(\vec{p}) v_{s}(\vec{p}) e^{i p x}\right],
\label{eq:decompD}
\ee
where $s=\pm 1/2$ is 
the projection of the particle's spin on a fixed spatial direction, 
$E_{\vec{p}}=p^0=+\sqrt{\vec{p}^2+m^2}$, 
$u_{s}(\vec{p})$ and $v_{s}(\vec{p})$ are the 
positive- and negative-energy solutions of the Dirac equation in the momentum 
space, and $b_{s}(\vec{p})$ and $d_{s}^\dag(\vec{p})$ are the annihilation 
operator for the particle and the creation operator for the antiparticle, 
respectively. The field $\psi$ thus annihilates the particle and creates its 
antiparticle, whereas the hermitian conjugate field annihilates the 
antiparticle and creates the particle. Because for Majorana fermions particle 
and antiparticle coincide, for them one has to identify $b_{s}(\vec{p})$ and 
$d_{s}(\vec{p})$, i.e.\ the Fourier expansion of Majorana fields takes the 
form%
\footnote{Expansions (\ref{eq:decompD}) and (\ref{eq:decompM}) are sometimes 
defined with a phase factor $\lambda$ in front of the creation operators. This 
factor, however, enters physical observables only together with other 
phase factors, discussed in sec.\ 15.2, i.e.\ it is not separately 
observable. We therefore choose $\lambda=1$ throughout this chapter.} 
\be
\psi(x)=\int\frac{d^3p}{(2\pi)^3\sqrt{2E_{\vec{p}}}}\sum_s \left[
b_{s}(\vec{p}) u_{s}(\vec{p}) e^{-i p x} + b^\dag_{s}
(\vec{p}) v_{s}(\vec{p}) e^{i p x}\right].
\label{eq:decompM}
\ee
It is possible (and convenient) to choose the phases of the spinors 
$u_s(\vec{p})$ and $v_s(\vec{p})$ in such a way that 
\be
v_s(\vec{p})={\cal C}\bar{u}_s^T(\vec{p})\,,\qquad 
u_s(\vec{p})={\cal C}\bar{v}_s^T(\vec{p})\,.
\label{eq:relat1}
\ee
{}From these relations it immediately follows that the field (\ref{eq:decompM}) 
satisfies the Majorana self-conjugacy condition (\ref{eq:C4}). The plane-wave 
decomposition of the Majorana fields (\ref{eq:decompM}) is reminiscent of the 
familiar Fourier expansion of 
the photon field $A_\mu(x)$, which also contains the creation and annihilation 
operators of only one kind, $a_\lambda(\vec{p})$ and $a_\lambda^\dag(\vec{p})$, 
because the photon 
is its own antiparticle. 

The action of the charge conjugation operation C  
amounts to interchanging the particle with its antiparticle without changing 
its momentum or spin polarization state. For a Dirac fermion field 
(\ref{eq:decompD}) it can therefore be represented as  
\be
{\rm C} b_s(\vec{p}) {\rm C}^{-1}= d_s(\vec{p})\,,\qquad
{\rm C} d_s^\dag(\vec{p}) {\rm C}^{-1}=b_s^\dag(\vec{p})\,.
\label{eq:chargeCon}
\ee
With the help of eq.~(\ref{eq:relat1}) one can readily make sure that 
applying to (\ref{eq:decompD}) the particle-antiparticle conjugation 
defined in eq.~(\ref{eq:C}) yields exactly the same result as the C conjugation 
(\ref{eq:chargeCon}). 
How about the Majorana fields? 
For them $d_s(\vec{p})=
b_s(\vec{p})$, so that the operation in eq.~(\ref{eq:chargeCon}) is just the 
trivial identity transformation which has no effect on the fields. 
The $\hat{C}$ operation also leaves the Majorana fields unchanged -- we have 
actually defined 
them through this condition, eq.~(\ref{eq:C4}). Thus, we conclude that for 
free massive 
fermion fields, both of Dirac and Majorana nature, the C and $\hat{C}$ 
conjugations are equivalent.  
As we already pointed out, the two operations are not equivalent when acting 
on chiral fields.

Consider now the equations of motion for the left-handed and 
right-handed components of a Majorana field. Eq.~(\ref{eq:M}) tells us 
that in the 4-component notation (\ref{eq:D5}) the lower 2-spinor is given 
by $\xi=-i\sigma^2\phi^*$. Substituting this into (\ref{eq:D6}) we 
find \cite{Case}
\bea
(\partial_0-\boldsymbol{\sigma}\cdot\boldsymbol{\nabla})\phi+m\sigma^2
\phi^*=0\,,
\label{eq:D6a}\\
(\partial_0+\boldsymbol{\sigma}\cdot\boldsymbol{\nabla})\sigma^2\phi^*-m
\phi=0\,.
\label{eq:D6b}
\eea
It is easy to see that the second of these equations is equivalent to the 
first one. Indeed, taking the complex conjugate of (\ref{eq:D6b}), multiplying 
on the left by $\sigma^2$ and using the relation $\sigma^2 
\boldsymbol{\sigma}^*\sigma^2 = -\boldsymbol{\sigma}$ we obtain 
 eq.~(\ref{eq:D6a}). 
Next, let us exclude $\phi^*$ from eqs.~(\ref{eq:D6a}) and (\ref{eq:D6b}).  
By acting on (\ref{eq:D6a}) with 
($\partial_0+\boldsymbol{\sigma}\cdot\boldsymbol{\nabla})$ and making use of 
(\ref{eq:D6b}) we find that $\phi$ satisfies the Klein-Gordon equation  
\be
(\partial^2+m^2)\phi=0\,.
\label{eq:KG}
\ee
This means that free Majorana particles obey the standard dispersion 
relation $E^2={\bf p}^2+m^2$. Thus, kinematically Dirac and Majorana 
fermions are indistinguishable. They can, however, in principle be 
told apart through their interactions, as we discuss below. 

Let us now 
turn to the Lagrangian of a free Majorana field. From eqs.~(\ref{eq:mass}) 
and (\ref{eq:psi1}) we find that the mass term in the Lagrangian is  
\be
{\cal L}_m=-\frac{m}{2}\left[\overline{(\psi_L)^c}\psi_L+
\overline{\psi_L} (\psi_L)^c\right]=\frac{m}{2}\left[\psi_L^T {\cal C}^{-1} 
\psi_L+ \overline{\psi_L}{\cal C}^{-1}  \overline{\psi_L}^T\right ]=
\frac{m}{2}\left[\psi_L^T {\cal C}^{-1} \psi_L+ h.c.\right] \,,
\label{eq:maj1}
\ee
where we have used the second equality in eq.~(\ref{eq:C2}), and the factor 
1/2 was introduced because ${\cal L}_m$ is quadratic in $\psi_L$. Thus, the 
Majorana Lagrangian can be written as 
\be
{\cal L}=\overline{\psi_L}\, i\gamma^\mu \partial_\mu \psi_L+ 
\frac{m}{2}\left[\psi_L^T {\cal C}^{-1} \psi_L+ h.c.\right]. 
\label{eq:maj1a}
\ee
Note that it is expressed solely in terms of $\psi_L$. In 
particular,  there is no kinetic term for the field $\psi_R$ because the 
left-handed and right-handed components of the Majorana field are not 
independent. The Lagrangian in (\ref{eq:maj1a}) can be cast in a more 
familiar form if we use the notation $\psi=\psi_L+(\psi_L)^c$. Then, up to a 
total derivative term that does not contribute to the action, the Lagrangian 
(\ref{eq:maj1a}) can be rewritten as%
\footnote{Indeed, 
$\bar{\psi}i\gamma^\mu \partial_\mu \psi=
\overline{\psi_L}\,i\gamma^\mu \partial_\mu \psi_L+
\overline{(\psi_L)^c}\,i\gamma^\mu \partial_\mu (\psi_L)^c$, and using 
eqs.~(\ref{eq:C}) and (\ref{eq:C1}) one can rewrite the last term as 
$\overline{(\psi_L)^c}\,i\gamma^\mu \partial_\mu (\psi_L)^c
=-\partial_\mu[\bar{\psi}_L\,i\gamma^\mu \psi_L]+
\bar{\psi}_L\,i\gamma^\mu \partial_\mu \psi_L$. 
Thus, we have 
$\bar{\psi}_L\,i\gamma^\mu \partial_\mu \psi_L=(1/2)
\bar{\psi}\,i\gamma^\mu \partial_\mu \psi$ + total derivative term. 
}
\be
{\cal L}=\frac{1}{2}\bar{\psi} i\gamma^\mu \partial_\mu \psi -
\frac{m}{2}\bar{\psi}\psi\,.
\label{eq:maj1b}
\ee

\vspace*{1.5mm}
\colorbox{lightgray}
{\parbox[t]{15cm}{\footnotesize 
It is not difficult to write down the Majorana Lagrangian 
in the 2-component notation. 
From $\psi_L=(\phi, 0)^T$ and (\ref{eq:maj1a}) we have 
\be
{\cal L}=\phi^\dag i(\partial_0-\boldsymbol{\sigma}\cdot\boldsymbol{\nabla})
\phi-\frac{1}{2}(\phi^T i\sigma^2 \phi + h.c.)\,.
\label{eq:maj1c}
\ee
By comparing the mass term in this expression with (\ref{eq:maj1}) one can see 
that in the 2-component formalism the role of the particle-antiparticle 
conjugation matrix ${\cal C}$ is played by $i\sigma^2$.  
}}

\vspace*{3.0mm}

{}From eq. (\ref{eq:maj1}) a very important difference between the Dirac and
Majorana mass terms follows. The Dirac mass terms $\bar{\psi}\psi$ are 
invariant with respect to the $U(1)$ transformations 
\be
\psi \to e^{i\alpha}\psi\,,\quad\quad
\bar{\psi} \to \bar{\psi}e^{-i\alpha}\,,
\label{eq:U1}
\ee
i.e.\ they conserve the 
charges associated with the corresponding transformations (electric charge, 
lepton or baryon number, etc.). At the same time, 
the Majorana mass terms have the structure $\psi_L \psi_L +h.c.$ and therefore 
they break all $U(1)$-charges by two units. Since the electric charge is 
exactly conserved, this in particular means that no charged particle can have 
Majorana mass. 

Another important point is that the Majorana mass term in eqs.~(\ref{eq:maj1}) 
and~(\ref{eq:maj1a}) do not vanish even though the matrix ${\cal C}^{-1}$ is 
antisymmetric (because so is ${\cal C}$). This follows from the fact that the 
fermionic quantum fields anticommute, and so 
the interchange of the two $\psi_L$ in $\psi_L^T{\cal C}^{-1}\psi_L$ yields 
an extra minus sign. Similar argument applies to the Majorana mass term in the 
2-component formalism in eq.~(\ref{eq:maj1c}) (note that the matrix $\sigma^2$ 
is antisymmetric). Thus, the Majorana mass is of essentially quantum nature.%
\footnote{Note, however, that 
formally one can also write the Majorana mass term at the classical 
level if one assumes that $\psi(x)$ is an anticommuting classical field, i.e.\ 
a field that takes as values Grassmann numbers.}

In the massless limit the difference between Dirac and Majorana particles 
disappears as both actually become Weyl particles. In particular, vanishing 
Majorana mass means that the free Lagrangian now conserves a $U(1)$ charge 
corresponding to the transformations (\ref{eq:U1}).   

Let us now briefly review the Feynman rules for Majorana particles
\cite{JLS,HabKane,GK,GZ,DEHK,Luty}. Unlike for a Dirac fermion, whose quantum 
field $\psi$ annihilates the particle and creates its antiparticle 
while $\psi^\dag$ annihilates the 
antiparticle and creates the particle, 
in the Majorana case the same field $\chi$ creates and 
annihilates the corresponding Majorana fermion. This leads to the existence 
of Wick contractions that are different from the standard ones. As a result, 
in addition to the usual Feynman propagator 
\be
S_F(x-x')\equiv \langle 0| T\chi(x)\bar{\chi}(x')|0 \rangle=
\int \frac{d^4 p}{(2\pi)^4}\frac{i(\slashed{p}+m)}{p^2-m^2+i\varepsilon}
e^{-i p (x-x')} 
\label{eq:propag1}
\ee
which coincides with the propagator of the Dirac fermion, 
there exist new types of propagators \cite{JLS,HabKane},  
\be
\langle 0| T \chi(x)\chi^T(x')|0\rangle=-S_F(x-x'){\cal C}~\quad
\mbox{and} \quad~ 
\langle 0| T \bar{\chi}^T(x)\bar{\chi}(x')|0\rangle={\cal C}^{-1}S_F(x-x')\,,
\label{eq:propag2}
\ee
where $\slashed{p}\equiv\gamma^\mu p_\mu$ and we have used the second equality 
in~(\ref{eq:C2}) and the Majorana condition $\chi^c=\chi$. 
Recall that Dirac fermions carry a conserved additive charge which is 
generically called the fermion number. The flow of this number is usually 
indicated on Feynman diagrams by arrows on the fermion lines which 
correspond to the standard propagator~(\ref{eq:propag1}). If a diagram 
contains a chain of fermion lines, the fermion number flow is continuous 
through this chain. 
As Majorana particles do not carry any conserved additive quantum number, 
there is no continuous flow of fermion number through Feynman diagrams  
in the Majorana case. This is reflected in the existence of the fermion number 
violating propagators (\ref{eq:propag2}), which can be graphically represented 
as lines with two arrows pointing in opposite directions (outwards for the 
first propagator in~(\ref{eq:propag2}) and inwards for the second one). 
In addition, each term of the interaction Lagrangian that contains Majorana 
fields gives rise to several vertices, depending on the direction of the 
arrows on the incoming and outgoing lines of Majorana particles. Some of these 
vertices also contain 
the particle-antiparticle conjugation matrix ${\cal C}$ \cite{JLS,HabKane}.
Special care should be taken to get the correct relative signs between 
different diagrams contributing coherently to the same amplitude. The rules 
are completed by requiring that diagrams with Majorana fermion loops 
have an extra factor 1/2 due to the permutation symmetry of the Majorana 
particles. 

The resulting Majorana Feynman rules are rather complicated. They can, however, 
be simplified by noting that the matrices ${\cal C}$ 
(or ${\cal C}^{-1}$)  
that are present in some vertices are always either canceled by the 
corresponding matrices in the propagators in eq.~(\ref{eq:propag2}) 
or eliminated through the proper attribution of the spinors 
to the external fermionic legs of the diagram with the help of 
eq.~(\ref{eq:relat1}). This leads to much simpler Majorana Feynman  rules, 
with propagators and vertices not containing explicitly the 
matrix ${\cal C}$ \cite{GK,GZ,DEHK}. In this case the Feynman rules 
include just the usual propagator~(\ref{eq:propag1})
for Majorana fermions, and the number of vertices corresponding to each term 
of the interaction Lagrangian is at most two. The Majorana fermion propagators 
are depicted by lines with no arrows. Instead of the fermion number flow 
(which is not conserved) the notion of a fermion flow is introduced. To each 
diagram a certain (but arbitrary) direction of the fermion flow is attributed,
which is used simply as a bookkeeping device; the analytic expressions for the 
amplitudes are independent of the chosen direction of this flow.

Finally, in appendix B of ref.~\cite{Luty} a very simple set of Majorana 
Feynman rules is suggested, based on the elimination of the adjoint Majorana 
fields $\bar{\chi}$ from the kinetic as well as the interaction terms of the 
Lagrangian through the relation $\bar{\chi}=\overline{\chi^c}=
-\chi^T{\cal C}^{-1}$.

For more detailed discussions of Majorana Feynman rules we refer the reader
to refs.~\cite{JLS,HabKane,GK,GZ,DEHK,Luty}.

\subsection*{\label{sec:discrete}15.2\, C, P, CP and CPT properties of Majorana 
fermions}
Since Majorana fermions are their own antiparticles, they are expected to 
have special properties with respect to C, CP and CPT transformations. 

Consider first the charge conjugation C.%
\footnote{Here we mostly follow ref.~\cite{Kayser1}, though some of our 
phase conventions are different.}
As was discussed above, for massive spinor fields this operation coincides 
with the particle-antiparticle conjugation $\hat{C}$ defined in 
eq.~(\ref{eq:C}). The latter, however, without loss of generality can be 
modified by introducing an arbitrary phase factor $\eta_{\rm C}^*$ on the 
right-hand side. That is, instead of eq.~(\ref{eq:C}) we can define
\be
\psi^c\equiv\eta_{\rm C}^* \,{\cal C}\bar{\psi}^T
=\eta_{\rm C}^* i\gamma^2 \psi^*\,.
\label{eq:Ca}
\ee
Indeed, this will not affect the evolution equation satisfied by 
$\psi^c(x)$ as well as the relation $(\psi^c)^c=\psi$. 
The charge conjugation transformation C (\ref{eq:chargeCon}) can be 
modified accordingly, so that the equivalence between the C and $\hat{C}$ 
conjugations is maintained:
\be
{\rm C}b_s(\vec{p}){\rm C}^{-1} = \eta_{\rm C}^* d_s(\vec{p})\,,\qquad\quad
{\rm C}d_s^\dag(\vec{p}){\rm C}^{-1} = \eta_{\rm C}^* b_s^\dag(\vec{p})\,.
\label{eq:chargeCon1}
\ee

Eqs.~(\ref{eq:Ca}) and~(\ref{eq:chargeCon1}) apply to arbitrary spin-1/2 
fermions; let us now discuss Majorana fields. Because for them 
${\cal C}\bar{\psi}^T=\psi$, the new definition of $\hat{C}$-conjugation 
(\ref{eq:Ca}) implies that the Majorana condition (\ref{eq:C4}) now takes the 
form 
\be
\psi^c(x)=\eta_{\rm C}^* \psi(x)\,.
\label{eq:C4a}
\ee
Since for Majorana fields one has to identify $d_s(\vec{p})=b_s(\vec{p})$, 
eq.~(\ref{eq:chargeCon1}) becomes  
\be
{\rm C}b_s(\vec{p}){\rm C}^{-1} = \eta_{\rm C}^* b_s(\vec{p})\,,\qquad\quad
{\rm C}b_s^\dag(\vec{p}){\rm C}^{-1} = \eta_{\rm C}^* b_s^\dag(\vec{p})\,.
\label{eq:chargeCon2}
\ee
Hermitian conjugation of the first of these two equalities 
yields ${\rm C}b_s^\dag(\vec{p}){\rm C}^{-1} = \eta_{\rm C} b_s^\dag(\vec{p})$. 
The consistency of this relation and the second one in 
eq.~(\ref{eq:chargeCon2}) requires that $\eta_C$ be real, i.e. $\eta_C=\pm 1$. 
Next, let us apply the second relation in~(\ref{eq:chargeCon2}) to the 
vacuum state. Assuming the vacuum to be even under the charge conjugation, we 
find  
\be
{\rm C} |\vec{p},s\rangle = \eta_{\rm C}|\vec{p},s\rangle\,,
\label{eq:C1b}
\ee
where $|\vec{p},s\rangle$ 
is the 1-particle Majorana state with momentum $\vec{p}$ and spin projection 
$s$.  

The Majorana condition~(\ref{eq:C4a})  
and eq.~(\ref{eq:C1b}) imply  
that the Majorana state is an eigenstate of charge 
conjugation C, and $\eta_C$ is its charge parity. It should be stressed, 
however, that 
this is, strictly speaking, only valid when C is exactly conserved. 
The above description certainly applies to free Majorana fermions, since 
the corresponding action is charge conjugation invariant.%
\footnote{This can be most easily seen if we rewrite the kinetic term 
of the Lagrangian ${\cal L}_k = \bar{\psi}i\gamma^\mu\partial_\mu\psi$ 
as ${\cal L}_k=(1/2)[\bar{\psi}i\gamma^\mu\partial_\mu
\psi-\partial_\mu \bar{\psi}\cdot i\gamma^\mu\psi]$+total derivative term 
and apply the $\hat{C}$-conjugation (which for massive fermion fields is 
equivalent to charge conjugation C) to the full Lagrangian of free 
Majorana particles. 
}
However, the charge parity $\eta_{\rm C}$, apart from being real, is completely 
arbitrary and therefore unphysical in this case.    
A physical (i.e.\ interacting) Majorana particle is an eigenstate 
of C only when all its interactions are C-invariant.%
\footnote{
An example of such a Majorana fermion is the photino -- the supersymmetric 
partner of the photon \cite{Nilles,HabKane,Drees,Feng1}.
} 
The C-parity of a Majorana particle is then constrained by the 
C-transformation properties of the other fields 
that enter its interaction Lagrangian. 
If C is only an approximate symmetry of the theory, 
the Majorana particle will be an approximate eigenstate of C, to the same 
extent to which charge conjugation invariance is satisfied. 

The situation is completely different for neutrinos, which are the prime  
candidates for being Majorana particles. The point is that their 
charged-current weak interactions are maximally C-violating.
Indeed, these interactions are left-handed (i.e.\ of the $V-A$ form), 
whereas charge conjugation would transform them into the 
right-handed ($V+A$) interactions which do not exist in the standard model 
based on the gauge group $SU(2)_L\times U(1)$. 
Thus, for Majorana neutrinos C-parity does not bear any physical sense. 

However, CP is a good approximate symmetry of the leptonic sector of the 
standard model. Indeed, it is is an exact symmetry of the gauge interactions, 
and in the minimally extended (to include non-zero neutrino mass) standard 
model it can only be violated by the neutrino mass generating sector. The 
corresponding CP-violation effects are very difficult to observe 
-- in particular, they have not been unambiguously observed by the time 
of publication of this book. Therefore, in many situations CP violation in the 
leptonic sector can be ignored. In other words, in some regards 
Majorana neutrinos can be considered as CP-eigenstates with certain 
CP-parities. This, however, is not in general true when possible CP-violating 
effects play a major role. We discuss these effects in secs.\
15.3, 15.7 and 15.8.

The properties of Majorana particles with respect to CP (assuming that it 
is a good symmetry) and CPT can be studied similarly to their properties 
under C transformation \cite{Kayser1}. The results are summarized in Table 1. 

\vspace*{1.5mm}
\colorbox{lightgray}
{\parbox[t]{15cm}{\footnotesize 
In deriving the properties of Majorana neutrinos under the discrete symmetries 
one can make use of the following properties of the spinors $u_s(\vec{p})$ and 
$v_s(\vec{p})$: 
\bea
&&
\gamma^0 u_s(\vec{p})=u_s(-\vec{p}),\qquad\qquad\qquad\qquad~~
\gamma^0 v_s(\vec{p})=-v_s(-\vec{p}),\nonumber \\
&&
u_s^*(\vec{p})=
(-1)^{s+1/2}
\gamma^1\gamma^3 
u_{-{}s}(-\vec{p}),\quad\quad~~
v_s^*(\vec{p})=
(-1)^{s+1/2}
\gamma^1\gamma^3 
v_{-{}s}(-\vec{p}),
\label{eq:propp1}
\eea
where the sign factors 
in the second line correspond to the phase convention 
$\gamma_5 u_s(\vec{p})=(-1)^{s-1/2}v_{-s}(\vec{p})$.
This choice of the phases is consistent with that in eq.~(\ref{eq:relat1}). 
It should also be kept in mind that, while C and P are unitary operators,
T is antiunitary, and so is CPT.  
}}

\begin{table}[h] 
\begin{tabular}{llll}
\hline\hline
Symmetry & ~~Effect on $\psi(t,\vec{x})$ &
~~Effect on $|\vec{p},s\rangle$ & ~~Restriction 
\vspace*{-1.2mm} \\
operation &{} &{} & \\
\hline 
C & ~~$\eta_{\rm C}^*i\gamma^2 \psi^*(t,\vec{x})$ & 
~~~$\eta_{\rm C}|\vec{p},s\rangle$ & ~~~$\eta_{\rm C}=\pm 1$ \vspace*{0.3mm}\\ 
CP & ~~$\eta_{\rm CP}^*i\gamma^0\gamma^2\psi^*(t,-\vec{x})$ & 
~~~$\eta_{\rm CP}|-\vec{p},s\rangle$ & ~~~$\eta_{\rm CP}=\pm i$
\vspace*{0.3mm}
\\ 
CPT & $-\eta_{\rm CPT}^*\gamma_5\psi^*(-t,-\vec{x})$ & 
~~~$\eta_{\rm CPT}^s|\vec{p},-s\rangle$ & ~~~$\eta_{\rm CPT}=\pm i$\\
\hline\hline
    \end{tabular}
    \caption{\label{tab:discrete} Effects of C, CP and CPT operations on 
a Majorana field $\psi(t,\vec{x})$ and on the corresponding one-particle 
Majorana state $|\vec{p},s\rangle$. Here $\eta_{\rm CPT}^s\equiv (-1)^{s-1/2}
\eta_{\rm CPT}$. }
\end{table}

\subsection*{\label{sec:osc}15.3\, Mixing and oscillations of Majorana 
neutrinos}

In sec.\ 15.1 we considered the mass term of a lone Majorana particle, 
eq.~(\ref{eq:maj1}). This 
expression is readily generalized to the case when there are 
$n$ Majorana fermions which in general can mix with each other: 
\be
{\cal L}_m=-\frac{1}{2}\left[\overline{(\psi_L)^c}\,m\, \psi_L+
\overline{\psi_L}\,m^\dagger\, (\psi_L)^c\right]=
\frac{1}{2}\left[\psi_L^T {\cal C}^{-1} m	
 \psi_L+ h.c.\right].
\label{eq:maj1aa}
\ee
Here $\psi=(\psi_1,...,\psi_n)^T$ 
and $m$ is an $n\times n$ matrix. Using the anticommutation property of the 
fermion fields and ${\cal C}^T=-{\cal C}$, it is easy to show that the matrix 
$m$ must be symmetric: $m^T=m$. 
Eq.~(\ref{eq:maj1aa}) applies to any set of Majorana particles; 
in the rest of this section we shall specifically consider Majorana neutrinos 
and their oscillations. 

\subsubsection*{15.3.1\, Neutrinos with a Majorana mass term}

Consider first the case 
of $n$ standard lepton generations, consisting each of an $SU(2)_L$-doublet 
of left-handed neutrino and charged lepton fields $l_{\alpha}=
(\nu_{\alpha L}, e_{\alpha L})^T$ 
and an $SU(2)_L$-singlet right-handed charged lepton field $e_{\alpha R}$ 
($\alpha=e,\mu,\tau,...)$ \cite{GribPont,Pont2}. 
In the standard model extended to include the mass 
generation mechanism for Majorana neutrinos, the terms of the Lagrangian 
that are relevant to neutrino oscillations include the charged-current (CC) 
weak interaction term and the mass terms of the charged leptons and neutrinos: 
\be
{\cal L}_{w+m}=
-\frac{g}{\sqrt{2}}(\bar{e}_{L}'\gamma^\mu
\,\nu_{L}')\,W_\mu -\bar{e}_{L}'m_l'\, e_{R}'+\frac{1}{2}
\nu_{L}'^{T}\,{\cal C}^{-1}m'\,\nu_{L}'~+~h.c.
\label{eq:L2}
\ee
Here $g$ is the CC gauge coupling constant, $W_\mu$ is the $W^-$-boson field, 
and all leptons are assembled in vectors in the 
generation space: $\nu_L'=(\nu_{eL},\, \nu_{\mu L},\, \nu_{\tau L},...)^T$ and 
similarly for $e_L'$ and $e_R'$.  
Since the CC weak interaction Lagrangian 
in eq.~(\ref{eq:L2}) is diagonal in the chosen basis, 
the latter is called the weak-eigenstate basis.  
The matrices $m_l'$ and $m'$ are, however, in general not diagonal in this 
basis. 
For $n$ leptonic generations, the mass matrix of the charged leptons $m'_l$ 
is a general complex $n\times n$ matrix, whereas the Majorana mass matrix of 
neutrinos $m'$ is a complex symmetric $n\times n$ matrix.
Recall now that an arbitrary square matrix $A$ can be diagonalized by a 
bi-unitary transformation according to $A_{diag}=V_1^\dag A V_2$, where 
$A_{diag}$ is a diagonal matrix with non-negative diagonal elements. 
Similarly, a symmetric square matrix $B$ is diagonalized by a transformation 
with a single unitary matrix: $B_{diag}=U^T B U$, where all the diagonal 
elements of $B_{diag}$ are non-negative. We therefore  
perform the basis transformations of the lepton fields according to 
\be
e_L'=V_L\,e_L\,,\quad e_R'=V_R\,e_R\,,\quad \nu_L'=U_L\,\nu_L\,,
\label{eq:unit1}
\ee
with $V_L$, $V_R$ and $U_L$ being unitary matrices chosen such that they 
diagonalize the charged-lepton and neutrino mass matrices:
\be
V_L^\dag m_l' V_R~=~m_l\,,\qquad U_L^T m' U_L~=~m\qquad 
(m_l,\, m ~-~ {\rm diagonal ~mass ~matrices})\,.
\label{eq:diag1}
\ee
Note that the kinetic Lagrangians of neutrinos and of 
the left- and right-handed charged leptons are invariant under these  
transformations. 
In the new (unprimed) 
basis eq.~(\ref{eq:L2}) takes the form 
\be
{\cal L}_{w+m}=-\frac{g}{\sqrt{2}}\sum_{\alpha,i}\bar{e}_{\alpha L}\gamma^\mu\,
U_{\alpha i}\,
\nu_{iL}\,W_\mu -\sum_\alpha m_{l\alpha} \bar{e}_{\alpha L}e_{\alpha R}+
\frac{1}{2} \sum_i m_{i} \nu_{iL}^T\,{\cal C}^{-1}\nu_{iL} + h.c.
\label{eq:L3}
\ee
Here $m_{l\alpha}$ ($a=e,\mu,\tau,...$) and $m_i$ ($i=1,2,3,..$) are the 
diagonal elements of the mass matrices $m_l$ and $m$ respectively, i.e.\ they 
are the masses of the charged leptons and of the mass-eigenstate neutrinos. 
The matrix 
\be
U\equiv 
V_L^\dag U_L
\label{eq:lmix}
\ee
is the leptonic mixing matrix, also called the Pontecorvo-Maki-Nakagawa-Sakata 
(PMNS) matrix \cite{Pont,MNS}.  
The flavour-eigenstate neutrino fields are defined as 
\be
\nu_{\alpha L}=\sum_{i=1}^n U_{\alpha i}\nu_{iL}\,.
\label{eq:mix1}
\ee
In terms of these fields the CC-interaction part of the Lagrangian 
(\ref{eq:L3}) takes the form ${\cal L}_{w}=-\frac{g}{\sqrt{2}}\sum_\alpha
\bar{e}_{\alpha L}\gamma^\mu\,\nu_{\alpha L}\,W_\mu +h.c.$, i.e.\ the flavour 
eigenstates $\nu_e$, $\nu_\mu$, $\nu_\tau$,... are the neutrinos emitted or 
absorbed together with the charged leptons $e,\mu,\tau$,... respectively. 

Let us introduce the 4-component neutrino fields
\be
\chi_i=\nu_{iL}+(\nu_{iL})^c\,,\qquad i=1,...,n\,.
\label{eq:chi}
\ee
Using the relations 
$\nu^T_{iL}\,{\cal C}^{-1}\nu_{iL}=-\overline{(\nu_{iL})^c}\, \nu_{iL}$, 
\,$(\nu^T_{iL\,}{\cal C}^{-1}\nu_{iL})^\dag=-\overline{\nu_{iL}}(\nu_{iL})^c$
which follow from eq.~(\ref{eq:C2}), one can rewrite the neutrino mass term 
in eq.~(\ref{eq:L3}) as  
\be
{\cal L}_{\nu m}=-\frac{1}{2}\sum_{i=1}^n 
m_i\bar{\chi}_i \chi_i\,.
\label{eq:mm}
\ee
Eqs.~(\ref{eq:mix1}) and (\ref{eq:chi}) mean that the 
neutrino flavour eigenstates $\nu_{\alpha L}$ are linear superpositions of the 
left-handed components of the $n$ mass eigenstates $\chi_i$. 
{}From eq.~(\ref{eq:chi}) it follows that 
$\chi_i^c=\chi_i$, i.e.\ the massive neutrino fields are Majorana fields 
in the case we consider. 

The Lagrangian (\ref{eq:L2}) (or equivalently (\ref{eq:L3})) implies that 
massive neutrinos are in general mixed and leads to the phenomenon of neutrino 
flavour oscillations \cite{Pont,MNS}.
The oscillation probability, i.e.\ the probability 
that a relativistic neutrino produced as a flavour eigenstate $\nu_\alpha$ 
will be in a flavour eigenstate $\nu_\beta$ after having propagated a distance 
$L$ in vacuum, is \cite{oscillTheory,oscTheory2}
\be
P(\nu_\alpha\to\nu_\beta; L) = \Big|\sum_i U_{\beta i}^{}\; 
e^{-i \frac{\Delta m_{ij}^2}{2p} L} \;U_{\alpha i}^*\Big|^2\,,
\label{eq:prob1}
\ee 
where $p$ is the modulus of the mean momentum of the neutrino state, 
$\Delta m_{ij}^2=m_i^2-m_j^2$ and 
for the index $j$ one can take any fixed value between 1 and $n$. 
Eq.~(\ref{eq:prob1}) can be equivalently written as 
\be
P(\nu_\alpha\to\nu_\beta; L) = \sum_{i,j} {\rm Re}
(U_{\alpha j}U_{\beta j}^*U_{\alpha i}^*U_{\beta i}^{})
\cos\Big(\frac{\Delta m_{ij}^2}{2p} L\Big)+\sum_{i,j}{\rm Im}
(U_{\alpha j}U_{\beta j}^*U_{\alpha i}^*U_{\beta i}^{})
\sin\Big(\frac{\Delta m_{ij}^2}{2p} L\Big).
\label{eq:prob2}
\ee 

Let us now discuss the general properties of the leptonic mixing matrix $U$.
Being an $n\times n$ unitary matrix, it depends on $n^2$ independent 
parameters, of which $n(n-1)/2$ are mixing angles 
and $n(n+1)/2$ are complex phases. Not all of these phases 
are physical, though.
As follows from eq.~(\ref{eq:L3}), one can always remove $n$ phases from 
$U$ by rephasing the left-handed fields of charged leptons according  
to $e_{\alpha L}\to e^{i\varphi_\alpha}e_{\alpha L}$, which allows one to 
fix the phases of one column of the matrix $U$. 
If one also rephases the right-handed charged 
lepton fields in the same way, i.e.\ $e_{\alpha R}\to e^{i\varphi_\alpha}
e_{\alpha R}$, the phase change of the fields $e_{\alpha L}$ in the mass term 
of the charged leptons in~(\ref{eq:L3}) will be compensated, and therefore 
eq.~(\ref{eq:L3}) will remain unchanged. 
It is easy to see that the kinetic terms of the Lagrangian of 
$e_{L}$ and $e_R$ are also invariant under the field rephasing. Thus, the 
leptonic Lagrangian is invariant with respect to the above rephasing of the 
charged lepton fields, which means that $n$ out of the $n(n+1)/2$ phases in 
$U$ are unphysical. 

How about rephasing the neutrino fields? Consider first the case when neutrinos 
are Dirac particles (which means that we should add $n$ right-handed neutrino 
fields to our model). Then their mass term is similar to that of the charged 
leptons. In that case it is possible to similarly rephase the left-handed and 
right-handed neutrino fields without modifying their mass term, which could 
be used to fix the phases of the elements of one line of the matrix $U$. 
However, the number of the phases that can be removed from $U$ in this way is 
$n-1$ rather than $n$, because the phase of one element which is at the 
intersection of the selected line of $U$ and the column whose phases have  
already been fixed by the rephasing of the charged lepton fields can no longer 
be modified. 
Thus, the total number of physical 
phases characterizing the leptonic matrix $U$ in the case of Dirac neutrinos is
\be
N_{ph}^{\rm D}=\frac{n(n+1)}{2}-n-(n-1)=\frac{(n-1)(n-2)}{2}\,.
\label{eq:Ndir}
\ee
Note that the quantities $U_{\alpha j}U_{\beta j}^*U_{\alpha i}^*
U_{\beta i}^{}$ that enter the expression (\ref{eq:prob2}) for the 
oscillation probabilities 
are invariant with respect to the rephasing of the charged-lepton and 
neutrino fields.

Let us now return to Majorana neutrinos. In this case the neutrino fields 
cannot be rephased since the Majorana mass term is of the type $\nu_L\nu_L
+h.c.$ rather than $\bar{\nu}_L\nu_R+h.c.$, and therefore it is not 
rephasing-invariant. As a result, the total number of physical phases 
characterizing the leptonic matrix $U$ in the case of the Majorana neutrinos is
\be
N_{ph}^{\rm M}=\frac{n(n+1)}{2}-n=\frac{n(n-1)}{2}\,.
\label{eq:Nmaj}
\ee
The extra $n-1$ physical phases that are present in the Majorana neutrino case 
can be collected in a diagonal matrix of phases which is factored out of 
$U$ as a right-hand side multiplier. Indeed, 
from eq.~(\ref{eq:L3}) it is seen that such a factorization isolates in 
a diagonal factor the phases that could have been absorbed into the 
rephasing of the $\nu_L$ fields if the neutrinos were Dirac particles. Thus, 
one can write 
\be
U=\tilde{U}\cdot diag(1, e^{i\varphi_1}, e^{i\varphi_2},\dots,  
e^{i\varphi_{n-1}})\equiv \tilde{U}K\,.
\label{eq:fact1}
\ee 
Here the matrix $\tilde{U}$ contains only $(n-1)(n-2)/2$ phases that are 
relevant also in the Dirac neutrino case, whereas the factor $K$ contains 
the extra `Majorana-type' phases. The position of the unit element in $K$ is 
irrelevant, as the overall phase of the matrix $U$ is unobservable. From 
eqs.~(\ref{eq:Ndir}) and~(\ref{eq:Nmaj}) it follows that in the Dirac neutrino 
case the mixing matrix $U$ contains physical complex phases only for $n\ge 3$ 
generations, whereas in the Majorana case the physical phases are in general 
there for $n\ge 2$.

The reason why we paid so much attention to the phases of the leptonic mixing 
matrix $U$ is that they lead to CP-violating effects in the 
leptonic sector. CP-conjugation transforms left-handed neutrinos into 
their antiparticles -- right-handed antineutrinos. Since CP-conjugation of a 
fermionic field includes complex conjugation (see Table~1), the oscillation 
probability $P(\bar{\nu}_a\to \bar{\nu}_b;L)$ is described by the right-hand 
side of eq.~(\ref{eq:prob1}) with the matrix $U$ substituted by $U^*$. 
Complex conjugation means that the signs of all the phases characterizing 
$U$ must be flipped; for this reason these phases are called CP-violating 
phases. In particular, if $U$ bears non-removable complex phases beyond 
those contained in the matrix $K$, neutrino oscillations violate CP-invariance, 
i.e.\
\be 
\Delta P_{\alpha\beta}^{\rm CP}(L)\equiv 
P(\nu_\alpha\to \nu_\beta;L)-P(\bar{\nu}_\alpha\to \bar{\nu}_\beta;L)\ne 0\,.
\label{eq:DeltaCP}
\ee
The quantity $\Delta P_{\alpha\beta}^{\rm CP}(L)$ coincides (up to the factor 
of two) with the CP-odd part of the $\nu_\alpha\to \nu_\beta$ oscillation 
probability, which is given by the second term on the right-hand side of 
eq.~(\ref{eq:prob2}), whereas the first term in (\ref{eq:prob2}) corresponds 
to the CP-even part of the probability. 

Can one find out whether neutrinos are Dirac or Majorana particles by studying 
neutrino oscillations? Unfortunately, this is not possible. It turns out that 
for Dirac neutrinos the oscillation probabilities are given by exactly the same 
formulas, eqs.~(\ref{eq:prob1}) or (\ref{eq:prob2}), as those for Majorana 
neutrinos. Moreover, the extra Majorana-type phases that enter the 
leptonic mixing matrix $U$ in the Majorana neutrino case are not observable 
in neutrino oscillations. Indeed, the matrix $K$ in 
eq.~(\ref{eq:fact1}) and the matrix $\exp[-i(\Delta m_{ij}^2/2p)L]$ for fixed 
$j$ that enters eq.~(\ref{eq:prob1}) are both diagonal and therefore 
commute, so that $K$ drops out of the expression for the oscillation 
probability (\ref{eq:prob1}). A similar argument applies to neutrino 
oscillations in matter. The Majorana-type phases, however, enter some 
other physical observables and so are in general observable 
quantities. We discuss these observables in secs.\ 15.7 and 15.8. 

For future reference, we give here the leptonic mixing matrix $U$ for the 
case of three leptonic generations with Majorana neutrinos in the so-called 
standard parameterization: 
\begin{align}
U & 
&=\left( \begin{matrix} c_{12} c_{13} &
 s_{12} c_{13} & s_{13} e^{-i \delta_{\rm CP}} \\ -s_{12}
 c_{23} - c_{12} s_{13} s_{23} e^{i \delta_{\rm CP}}
& c_{12} c_{23} - s_{12} s_{13} s_{23} e^{i \delta_{\rm CP}} & c_{13} s_{23}~~ 
\\ s_{12} s_{23} - c_{12} s_{13} c_{23} e^{i \delta_{\rm CP}} &
-c_{12} s_{23} - s_{12} s_{13} c_{23} e^{i \delta_{\rm CP}} & c_{13}
c_{23}~~ \end{matrix} 
\right)\!
\left(
\begin{matrix}1 & 0 & 0\\
0 & e^{i\varphi_1} & 0\\
0 & 0&  e^{i\varphi_2}
\end{matrix}
\right).
\label{eq:U2}
\end{align}
Here $c_{ij}\equiv\cos\theta_{ij}$, $s_{ij}\equiv\sin\theta_{ij}$, where 
$\theta_{ij}$ are the mixing angles, $\delta_{\rm CP}$ is the Dirac-type 
CP-violating phase and $\varphi_{1,2}$ 
are the Majorana-type CP-violating phases.

\subsubsection*{15.3.2\, General case of Dirac + Majorana mass term
\label{eq:D+M}
}

Consider now the case in which, in addition to $n$ standard leptonic 
generations with left-handed neutrino fields $\nu_{\alpha L}$ being parts 
of $SU(2)_L$ leptonic doublets, there exist $k$ right-handed neutrino fields 
$\nu_{\sigma R}$ that are electroweak singlets, i.e.\ singlets with respect to 
both weak isospin group $SU(2)_L$ and hypercharge $U(1)$ 
\cite{BP1,Barg,Bil1,Cheng,SchVal,Kobz}. Such neutrinos don't have 
electroweak gauge interactions (though may have, e.g., Yukawa interactions 
with leptonic and Higgs doublets) and therefore are often called `sterile 
neutrinos'.  In contrast to this, the usual $SU(2)_L$-doublet neutrinos 
$\nu_L$ are called `active neutrinos'.  

\vspace*{1.5mm}
\colorbox{lightgray}
{\parbox[t]{15cm}{\footnotesize 
Since $\nu_R$ are electroweak singlets, they do not contribute to the 
so-called chiral gauge anomalies and so their number is not fixed by the 
requirement of the anomaly cancellation. In particular, their number need 
not coincide with the number of the leptonic generations $n$, i.e.\ in 
general $k\ne n$. 
Let us stress that these extra neutrinos are sterile not because they are 
right-handed -- their $\hat{C}$-conjugates are left-handed and yet also sterile 
-- but because they are electroweak singlets.     
}}

\vspace*{1.5mm}

\noindent
In the considered case the most general neutrino mass term contains the 
Majorana masses $m_L$ and $m_R$ for the left-handed and right-handed 
neutrino fields respectively, 
as well as the Dirac mass $m_D$ that couples 
the $\nu_L$'s with the $\nu_R$'s: 
\be
{\cal L}_m=\frac{1}{2} \nu_{L}'^{T}\, {\cal C}^{-1}\, m_L\, \nu_{L}'-
\overline{\nu_{R}'}\,m_D\, \nu_{L}'+ \frac{1}{2}\nu_{R}'^T\, {\cal C}^{-1} 
\,m_R^*\,\nu_{R}' + h.c. 
\label{eq:Lm1}
\ee

\noindent
Here we have assembled the $n$ left-handed and $k$ right-handed neutrino 
fields into the vectors $\nu_L'$ and $\nu_R'$.  
The quantities $m_L$ and $m_R$ are complex symmetric $n\times n$ and 
$k\times k$ matrices respectively, $m_D$ is a complex $k\times n$ matrix, 
and we have introduced the right-handed Majorana mass matrix through its 
complex conjugate to simplify the further notation. 
Introducing the vector of $n+k$ left-handed fields  
\be
n_L=\left(\begin{array}{l}\nu_L'\! \\ \!(\nu_R')^{c}\! \end{array}\right)
=\left(\begin{array}{l}\nu_L'\\ \nu'^{c}_{~L} \end{array}\right),
\label{eq:n}
\ee
we can rewrite eq.~(\ref{eq:Lm1}) as  
\be
{\cal L}_m= \frac{1}{2}\, n_L^T \,{\cal C}^{-1} {\cal M}\, n_L +h.c. \,,
\label{eq:Lm1a}
\ee
where the matrix ${\cal M}$ has the form
\be 
{\cal M}=\left(\begin{array}{cc} m_L & m_D^T \\ m_D & 
m_R \end{array} \right). 
\label{eq:Lm1b} 
\ee 
In deriving eq. (\ref{eq:Lm1a}) we have used the relations 
\be 
(\psi_R^{T}\,{\cal C}^{-1} m^*\,\psi_R)^\dagger=(\psi^c_{\,L})^{T}\,
{\cal C}^{-1} m\,\psi^{c}_{\,L}\,,\quad 
\overline{\psi_R}\, m\, \psi_L=-(\psi^{c}_{\,L})^T\,{\cal C}^{-1}\,m\,\psi_L=
-\psi_L^{T}\,{\cal C}^{-1}\,m^T\, \psi^{c}_{\,L}\,, 
\label{eq:rel2} 
\ee 
which follow from eqs.~(\ref{eq:relat1}) and~(\ref{eq:C1}). 
The matrix ${\cal M}$ is complex symmetric, so it can be diagonalized 
with a single unitary matrix. We therefore write
\be
n_{aL}=\sum_{i=1}^{n+k}{\cal U}_{ai}\chi_{iL}\,,\qquad\quad
{\cal U}^T {\cal M}{\cal U}={\cal M}_d\,,
\label{eq:n1}
\ee
where ${\cal M}_d$ is a diagonal $(m+n)\times (m+n)$ matrix with non-negative 
diagonal elements ${\cal M}_{di}$. In terms of the fields $\chi_{iL}$ the 
neutrino mass term (\ref{eq:Lm1a}) reads
\be
{\cal L}_m= \frac{1}{2}\, \sum_{i=1}^{n+k}{\cal M}_{di}\, \chi_{iL}^T \,
{\cal C}^{-1}\chi_{iL} +h.c. \,.
\label{eq:Lm1c}
\ee
Introducing the 4-component massive neutrino fields $\chi_i$ as
\be
\chi_i=\chi_{iL}+(\chi_{iL})^c\,,\qquad i=1,\dots,n+k\,,
\label{eq:chi2}
\ee 
we can rewrite the neutrino mass term in the mass eigenstate basis 
(\ref{eq:Lm1c}) in the standard form:  
\be
{\cal L}_{m}=-\frac{1}{2}\sum_{i=1}^{n+k} 
m_i \bar{\chi}_i \chi_i\,.
\label{eq:mm1}
\ee

In eq.~(\ref{eq:n1}) the index $a$ can take $n+k$ values; we will denote 
collectively the first $n$ of them with $\alpha$ or $\beta$ and the last $k$ 
with $\sigma$ or $\rho$. Eq.~(\ref{eq:n1}) yields 
\be   
\nu_{\alpha L}=\sum_{i=1}^{n+k}{\cal U}_{\alpha i}\chi_{iL}\,,\qquad\quad
(\nu_{\sigma R})^c=\sum_{i=1}^{n+k}{\cal U}_{\sigma i}\chi_{iL}\,.
\label{eq:n2}
\ee
This means that left-handed active neutrinos and left-handed sterile 
antineutrinos are  linear combinations of the left-handed components of all 
$n+k$ mass-eigenstate fields $\chi_i$. 

{}From eq.~(\ref{eq:chi2}) it follows that $\chi_i^c=\chi_i$, i.e.\ the 
neutrino mass eigenstates are Majorana fermions in the case we currently 
consider, just as in the pure Majorana mass term case discussed in the 
previous subsection. This is a general feature of fermion mass models: if the 
fermions possess Majorana mass terms, then, independently of whether or not 
the Dirac mass terms are also present, the mass eigenstates are always Majorana 
particles. 

\vspace*{1.5mm}
\colorbox{lightgray}
{\parbox[t]{15cm}{\footnotesize 
This is actually easy to understand by counting the number of the field degrees 
of freedom. In the Majorana mass case studied in sec.\ 15.3.1 one has $n$ 
two-component neutrino fields, and the neutrino mass matrix has in general $n$ 
distinct eigenvalues. Each massive neutrino field then has two degrees of 
freedom, i.e.\ it should be a Majorana field. In the pure Dirac case there 
are $2n$ two-component fields ($n$ left-handed and $n$ right-handed), and the 
mass matrix has $n$ eigenvalues. This means that each mass eigenstate has 
four degrees of freedom, i.e.\ is a Dirac field. In the Dirac + Majorana 
mass case there are $n+k$ 2-component fields, $n$ left-handed and $k$ 
right-handed. The matrix ${\cal M}$ has $n+k$ in general distinct eigenvalues,
which means that each neutrino mass eigenstate is characterized by two degrees 
of freedom, i.e.\ is a Majorana field. 

\hspace*{3mm}
If some of the mass eigenvalues coincide, the corresponding 2-component 
Majorana fields can merge into 4-component Dirac ones. We will consider this 
phenomenon in the next subsection.
}}

\vspace*{1.5mm}

Let us now discuss neutrino oscillations in the Dirac + Majorana (D+M) 
mass scheme that we are now considering. Unlike in the pure Dirac or pure 
Majorana mass cases, in the 
D+M scheme two new types of neutrino oscillations become possible: 
active - sterile and sterile - sterile oscillations. The oscillations 
between the active neutrino species are described by the same 
expression as in eq.~(\ref{eq:prob1}) but with the matrix $U$ replaced by 
${\cal U}$ and the summation over $i$ extending from 1 to $n+k$. The 
probability of oscillations between the 
active and sterile neutrinos is 
\be 
P(\nu_{\alpha L}\to\nu^c_{\;\sigma L};L) = \left|\sum_{i=1}^{n+k} {\cal 
U}_{\sigma i}^{}\; e^{-i \frac{\Delta m_{ij}^2}{2p} L} \;{\cal 
U}_{\alpha i}^*\right|^2. 
\label{eq:prob3} 
\ee 
If a mechanism by which sterile neutrinos can be produced and detected exists,%
\footnote{Recall that even though sterile neutrinos don't have gauge 
interactions in the standard model, they may possess other interactions, such 
as the Yukawa ones, or extra gauge interactions in the extensions of the 
standard model (e.g., $SU(2)_R$ gauge interactions in left-right symmetric 
models).}  
one can in principle observe 
sterile - sterile neutrino oscillations, whose probability is 
\be 
P(\nu^c_{\,\sigma L}\to\nu^c_{\;\rho L};L) = \left|\sum_{i=1}^{n+k}
{\cal U}_{\rho i}^{}\; e^{-i \frac{\Delta m_{ij}^2}{2p} L} \;
{\cal U}_{\sigma i}^*\right|^2. 
\label{eq:prob4} 
\ee 
Eq.~(\ref{eq:prob4}) describes the oscillations between the left-handed 
sterile neutrino states $\nu^c_{\,L\sigma}=(\nu_{R\sigma})^c$ and 
$\nu^c_{\,L\rho}=(\nu_{R\rho})^c$; the oscillations between the corresponding 
right-handed states $\nu_{R\sigma}$ and $\nu_{R\rho}$ can be obtained from 
eq.~(\ref{eq:prob4}) by replacing ${\cal U}\leftrightarrow {\cal U}^*$. 

If the sterile neutrinos are completely undetectable, one can only observe 
active - active and active - sterile oscillations, the latter manifesting 
themselves through disappearance of the active neutrinos.   

\subsubsection*{15.3.3\, Dirac and pseudo-Dirac neutrino limits in the D+M case}

As we have pointed out above, if Majorana mass terms are present in a fermion 
mass model, the mass-eigenstate fermions are always Majorana particles, even 
when the Dirac mass terms are present as well. One can expect, however, 
that in the limit when the Majorana mass terms are much smaller than the 
Dirac ones, the properties of the mass eigenstates would become close to 
those of Dirac fermions. To see how this happens, 
it is instructive to consider the 
simple one-generation neutrino case with Majorana and Dirac mass terms.
The quantities $m_L$, $m_R$ and $m_D$ are then just 
numbers, and ${\cal M}$ is a $2\times 2$ matrix. For simplicity we shall 
assume all the mass parameters to be real. This, in particular, means that 
CP is conserved in the neutrino mass sector, i.e.\ the free mass eigenstates 
are also eigenstates of CP. 
The matrix ${\cal M}$ can be
diagonalized by the transformation 
$O^T{\cal M} O={\cal M}_d$ where $O$ is a real orthogonal $2\times 2$ matrix 
and ${\cal M}_d=diag(m_1,\, m_2)$. 
We introduce the fields $\chi_L$ through $n_L=O\chi_L$, or 
\be
n_L=\left(\begin{array}{c}
\nu_L \\
\nu^c_{\,L}   
\end{array} \right)=
\left(\begin{array}{cc}
\cos\theta      & \sin\theta      \\
-\sin\theta     & \cos\theta    
\end{array} \right)
\left(\begin{array}{c}
\chi_{1L} \\
\chi_{2L}   
\end{array} \right)\,. 
\label{eq:mL2}
\ee
Here $\chi_{1L}$ and $\chi_{2L}$ are the left-handed components of the 
neutrino mass eigenstates.  The mixing angle $\theta$ is given by  
\be
\tan 2\theta=\frac{2 m_D}{m_R-m_L}\,,
\label{eq:theta}
\ee
and the neutrino mass eigenvalues are 
\be
m_{1,2}=\frac{m_R+m_L}{2}\mp \sqrt{\left(\frac{m_R-m_L}{2}\right)^2
+m_D^2}\,.
\label{eq:m1m2}
\ee
They are real but can be of either sign. The neutrino mass term can now be 
rewritten as 
\begin{eqnarray}
{\cal L}_m &=& \frac{1}{2}\, n_L^T \,{\cal C}^{-1} {\cal M}\, n_L+h.c.= 
\frac{1}{2}\,\chi_L^T \,{\cal C}^{-1} {\cal M}_d\, \chi_L +h.c. \nonumber \\ 
&=& \frac{1}{2}(m_1\, \chi_{1L}^T \,{\cal C}^{-1}\chi_{1L}+m_2\, 
\chi_{2L}^T \,{\cal C}^{-1}\chi_{2L})+h.c. =\frac{1}{2} (\, |m_1|\,
\overline{\chi}_1\chi_{1}+|m_2|\, \overline{\chi}_2\chi_{2}\,)\,. 
\label{eq:mL3}
\end{eqnarray}
Here we have defined 
\be
\chi_1=\chi_{1L}+\eta_1(\chi_{1L})^c\,,\quad\quad
\chi_2=\chi_{2L}+\eta_2(\chi_{2L})^c\,. 
\label{eq:chi1}
\ee
with $\eta_{i}=1$ or $-1$ for $m_{i}>0$ or $<0$ respectively. It follows 
immediately from eq.~(\ref{eq:chi1}) that the mass-eigenstates fields 
$\chi_1$ and $\chi_2$ describe Majorana neutrinos. The relative signs of the 
mass eigenvalues ($\eta_1$ and $\eta_2$) determine the relative CP parities of 
$\chi_1$ and $\chi_2$; the physical masses $|m_1|$ and $|m_2|$ are positive, as 
they should be. 

\vspace*{1.5mm}
\colorbox{lightgray}
{\parbox[t]{15cm}{\footnotesize 
Instead of using a real orthogonal matrix $O$ to diagonalize ${\cal M}$ one 
could employ a unitary matrix $U=O K$ with $K$ being a diagonal matrix of 
phases, as in eq.~(\ref{eq:fact1}). Choosing for the positive mass eigenvalues 
of ${\cal M}$ the diagonal elements of $K$ to be $1$ and for the negative 
ones $\pm i$, one can write $U^T {\cal M} U={\cal M}_d$, where ${\cal M}_d$ 
now does not have any negative diagonal elements. 
Correspondingly, eq.~(\ref{eq:mL2}) 
should be replaced with $n_L=U\chi_L$. In this way it is no longer  
necessary to introduce the factors $\eta_{1,2}$, 
i.e.\ instead of eq.~(\ref{eq:chi1}) we have 
\be
\chi_1=\chi_{1L}+(\chi_{1L})^c, \qquad \chi_2=\chi_{2L}+(\chi_{2L})^c.  
\label{eq:chi1b}
\ee
That the neutrino mass eigenstates corresponding to opposite signs of 
the mass parameters $m_1$ and $m_2$ defined in eq.~(\ref{eq:m1m2}) have 
opposite CP-parities is now a consequence of the fact that the matrix $U$ has 
one complex column. 
Indeed, let $m_1$ defined in (\ref{eq:m1m2}) be negative and $m_2$ positive.
Then from $n_L=U\chi_L=OK\chi_L$ and eq.~(\ref{eq:chi1b}) 
we have 
\be
\chi_1=\mp i\{c[\nu_L-(\nu_L)^c]+s[\nu_R-(\nu_R)^c]\},\qquad
\chi_2=s[\nu_L+(\nu_L)^c]+c[\nu_R+(\nu_R)^c].
\label{eq:new} 
\ee
Making use of the definition of the CP conjugation given in sec.\ 15.2 
and taking into account that it is described by a linear operator, 
one can readily make sure that the CP-parities of $\chi_1$ and $\chi_2$ are 
opposite. 
}}
\vspace*{1.5mm}

It is instructive to consider some limiting cases. In the limit of no
Majorana masses ($m_L=m_R=0$), pure Dirac case has to be recovered. Let 
us see how this limit can be obtained from the general D+M formalism.  
For $m_L=m_R=0$ the mass matrix (\ref{eq:Lm1b}) takes the form 
\be
{\cal M}=\left(\begin{array}{cc}
0      & m     \\
m      & 0    
\end{array} \right).
\label{eq:mL4}
\ee
This matrix corresponds to a conserved lepton number $L_{\nu_L}-
L_{\nu^c_{\,L}}=L_{\nu_L}+L_{\nu_{R}}$ which can be identified with the 
total lepton number $L$. Thus, the lepton number is conserved in this 
limiting case, as expected. Let us now check that the usual Dirac mass
term is recovered. 

The matrix ${\cal M}$ in (\ref{eq:mL4}) is diagonalized by the rotation 
(\ref{eq:mL2}) with $\theta=45^\circ$, and its eigenvalues are $-m$ and $m$. 
This means that we have two Majorana neutrinos that have the same mass, 
opposite CP parities and are maximally mixed. Let us demonstrate
that this is equivalent to having one Dirac neutrino of mass $m$. We have 
$\eta_2=-\eta_1=1$; from eqs.~(\ref{eq:mL2}) and (\ref{eq:chi1}) it then 
follows $\chi_1+\chi_2=\sqrt{2}(\nu_L+\nu_R)$, 
$\chi_1-\chi_2=-\sqrt{2}(\nu^c_{\,L}+\nu^c_{\,R})=-(\chi_1+\chi_2)^c$. 
This gives 
\be
\frac{1}{2}\,m\,(\overline{\chi}_1\chi_1 +\overline{\chi}_2\chi_2)=
\frac{1}{4}\,m\,[\overline{(\chi_1+\chi_2)}(\chi_1+\chi_2) +
[\overline{(\chi_1-\chi_2)}(\chi_1-\chi_2)]=m\,
\bar{\nu}_D \nu_D\,,
\label{eq:check1}
\ee
where
\be
\nu_D\equiv \nu_L +\nu_R\,.
\ee
The counting of the degrees of freedom also shows that we must have a Dirac 
neutrino in this case -- there are four degrees of freedom and just one 
physical mass. 
Thus, two maximally mixed degenerate Majorana neutrinos of opposite CP 
parities merge to form a Dirac neutrino. In sec.\ 15.7 we shall 
discuss neutrinoless double beta ($0\nu\beta\beta$) decay and show that 
this process can only take place if neutrinos are Majorana particles. 
We shall demonstrate there that in the limit when two 
degenerate in mass Majorana neutrinos merge into a Dirac neutrino, their 
contributions to the amplitude of the $0\nu\beta\beta$ decay exactly cancel, 
as they should. 

If the Majorana mass parameters $m_L$ and $m_R$ do not vanish but are small 
compared to $m_D$, the resulting pair of Majorana neutrinos will be 
quasi-degenerate with almost maximal mixing and opposite CP parities. The 
physical neutrino masses in this case are $|m_{1,2}|\simeq m_D \mp 
(m_L+m_R)/2 \simeq m_D$. Such a pair in many respects behaves as a Dirac 
neutrino and therefore sometimes is called a `pseudo-Dirac' (or a 
`quasi-Dirac') neutrino. In particular, its contribution to the 
$0\nu\beta\beta$ decay amplitude is proportional to the mass difference 
$|m_2|-|m_1|\simeq (m_L+m_R)$ which is much smaller than the mass of each 
component of the pair.

\subsection*{\label{sec:ssmech}15.4\, The seesaw mechanism of the neutrino mass 
generation }

The seesaw mechanism \cite{seesaw} provides a very simple and attractive 
explanation of the smallness of neutrino mass  
by relating it with the existence of a very large mass scale. 
In the simplest and most popular version of this mechanism, 
the requisite large mass scale is given by the masses of 
heavy electroweak-singlet Majorana neutrinos. 
Although the seesaw 
mechanism is most natural in the framework of the grand unified theories 
(such as $SO(10)$) or left-right symmetric models \cite{Mohapatra}, it also 
operates in the standard model extended to include the right-handed 
sterile neutrinos $\nu_R$. Indeed, as soon as the $\nu_R$'s are introduced, 
one can add to the Lagrangian of the model the Majorana mass term 
$(1/2)\nu_R^T\,{\cal C}^{-1}m_R\nu_R+h.c.$, which is allowed because 
$\nu_R$ are electroweak singlets. The Yukawa couplings of the right-handed 
neutrinos with the lepton doublets and the Higgs boson are also allowed, 
and after the spontaneous breaking of the electroweak symmetry they give rise 
to the Dirac mass term connecting the active and sterile neutrinos, 
similar to those that are present 
for the quarks and charged leptons. 
The resulting neutrino mass scheme is just the 
D+M one discussed in secs.\ 15.3.2 and 15.3.3, see eqs.~(\ref{eq:Lm1}) -
(\ref{eq:Lm1b}).   
Notice that in the standard model there is no Majorana mass term for 
left-handed neutrinos, i.e.\ $m_L=0$; however, 
$m_L$ is different from zero in some extensions of the standard model, 
so we shall keep it for generality.  

Because the right-handed neutrinos $\nu_R$ are electroweak singlets, 
the scale of their Majorana mass term need not be related to the electroweak 
scale. In particular, $m_R$ can be very large, possibly even at the Planck 
scale $M_{\rm Pl}\simeq 1.2\times 10^{19}$ GeV, grand unification scale or at 
some intermediate scale $M_I\sim 10^{10} - 10^{12}$ GeV which may be relevant 
for the physics of parity breaking. The seesaw suppression of the masses of 
active neutrinos is realized just in this case of a very large $m_R$. 
We therefore consider the limit 
in which the characteristic scales of the Dirac and Majorana neutrino masses 
satisfy  
\be
m_L, m_D\ll m_R\,.
\label{eq:ss}
\ee
Let us first discuss the one-generation case studied in sec.\ 15.3.3. 
The diagonalization of the mass matrix is 
then performed by the simple 
rotation (\ref{eq:mL2}). A straightforward calculation gives for the 
rotation angle $\theta$ and eigenvalues of the mass matrix ${\cal M}$ 
\be
\theta\simeq \frac{m_D}{m_R}\ll 1\,, \quad \quad  
m_1\simeq m_L-\frac{m_D^2}{m_R}\,,\quad \quad m_2 \simeq m_R\,,
\label{eq:ss1}
\ee
while the mass eigenstates are given by 
\be 
\chi_1\simeq \nu_L+\eta_1(\nu_L)^c \,, \quad\quad
\chi_2\simeq (\nu_R)^c +\eta_2 \nu_R \,.
\label{eq:ss2}
\ee
Thus, we have a very light Majorana mass eigenstate $\chi_1$ predominantly 
composed of the active neutrino $\nu_L$ and its $\hat{C}$-conjugate 
$(\nu_L)^c$, and a heavy eigenstate $\chi_2$, mainly composed of the sterile 
$\nu_R$ and $(\nu_R)^c$. The admixture of the sterile neutrino state $\nu_R$ 
in $\chi_1$ and that of the usual active neutrinos $\nu_L$ in $\chi_2$ are of 
the order of $m_D/m_R\ll 1$. As follows from eq.~(\ref{eq:ss1}), for 
$m_L\lesssim m_D^2/m_R$ it is 
the sterile neutrino 
being heavy that makes the usual active one light 
(which explains the name `the seesaw mechanism'). 

Consider now the 
case of $n$ standard generations of left-handed leptons and $k$ sterile 
neutrinos $\nu_R$. This is actually the case discussed in sec.\ 15.3.2, but 
now we want to specifically consider the limit of very high $\nu_R$ mass scale. 
Let us first decouple the light and heavy neutrino degrees of freedom. To this 
end, we block-diagonalize the matrix ${\cal M}$ in eqs.~(\ref{eq:Lm1a}), 
(\ref{eq:Lm1b}) according to 
\be
n_L=V \chi_L'\,,\quad\quad
V^T {\cal M}\,V \,= \,
V^T \left(\begin{array}{cc}
m_L & m_D^T  \\
m_D & M_R    
\end{array}\right)V\,=\,
 \left(\begin{array}{cc}
\tilde{m}_L & 0  \\
0           & \tilde{M}_R    
\end{array}\right),
\label{eq:ss3}
\ee
where $V$ is a unitary $(n+k)\times(n+k)$ matrix, 
$\tilde{m}_L$ and $\tilde{M}_R$ are symmetric $n\times n$ and $k\times k$ 
matrices, respectively, and we have changed the notation $m_R \to M_R$. Note 
that the fields $\chi'$ that block-diagonalize 
${\cal M}$ 
are not the fields of mass-eigenstate neutrinos, since 
the matrices $\tilde{m}_L$ and $\tilde{M}_R$ are not in general diagonal. 
They can be diagonalized by 
further unitary transformations. 
Correspondingly, $V$ is not the leptonic mixing matrix. 

We shall be looking for the matrix $V$ in the form \cite{grimlav} 
\be
V=
\left(\begin{array}{cc}
\sqrt{1-\rho\rho^\dag}  &  \rho \\
-\rho^\dag & \sqrt{1-\rho^\dag\rho}    
\end{array}\right),
\label{eq:V}
\ee
where $\rho$ is an $n\times k$ matrix. 
Note that $V$ is unitary by construction. Treating $\rho$ as perturbation 
and performing block-diagonalization of ${\cal M}$ approximately, we find  
\be
\rho^*\simeq m_D^T M_R^{-1}\,,\quad\quad 
\tilde{M}_R\simeq M_R\,,
\label{eq:ss4}
\ee
\be
\tilde{m}_L \simeq m_L - m_D^T M_R^{-1} m_D\,,\quad\quad ~
\label{eq:ss5}
\ee
These relations generalize those of eq. (\ref{eq:ss1}) to the case of $n$ 
active and $k$ sterile neutrinos. The diagonalization of the effective mass 
matrix $\tilde{m}_L$ then yields $n$ light Majorana neutrino fields which are 
predominantly composed of the fields of the usual (active) neutrinos $\nu_L$ 
and their $\hat{C}$-conjugates $(\nu_L)^c$, with very small ($\sim m_D/M_R$)
admixture of sterile neutrinos $\nu_R$; the diagonalization of $\tilde{M}_R$ 
produces $k$ heavy Majorana neutrino fields which are mainly composed of 
$\nu_R$ and $(\nu_R)^c$. 
This, in particular, means that the oscillations between the active and sterile 
neutrinos are suppressed in this case. 

\begin{figure}[h]
\begin{center}
\begin{picture}(359,80)(-5,-60)
\ArrowLine(40,-40)(100,-40)
\DashLine(100,-40)(100,5){5}
\Text(100,15)[c]{$\langle H\rangle$}
\Text(60,-50)[l]{\small $\nu_L$}
\Text(100,-50)[c]{\small $m_D$}
\ArrowLine(100,-40)(160,-40)
\Text(130,-50)[l]{\small $\nu_R$}
\Text(160,-40)[c]{$\times$}
\Text(160,-50)[c]{\small $M_R$}
\ArrowLine(220,-40)(160,-40)
\Text(190,-50)[l]{\small $\nu_R$}
\DashLine(220,-40)(220,5){5}
\Text(220,15)[c]{$\langle H\rangle$}
\Text(220,-50)[c]{\small $m_D$}
\ArrowLine(280,-40)(220,-40)
\Text(250,-50)[l]{\small $\nu_L$}
\end{picture}
\end{center}
\vspace{-0.6cm}
\caption{\small Seesaw mechanism of $\tilde{m}_L$ generation. The vacuum 
expectation values of the Higgs field are denoted by $\langle H\rangle$.  
}
\label{eq:seesawdiag}
\end{figure}
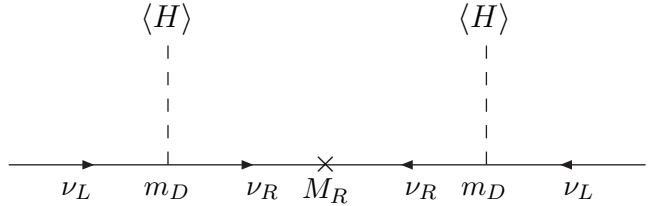

It is important that the active neutrinos get 
Majorana masses $\tilde{m}_L$ even if they have no `direct' masses, i.e.\ 
$m_L=0$, as it is the case in the standard model. The masses of the active 
neutrinos are then of the 
order of $m_D^2/M_R$. Generation of the effective Majorana mass of light 
neutrinos is diagrammatically illustrated in fig.~\ref{eq:seesawdiag}.

\vspace*{1.5mm}
\colorbox{lightgray}
{\parbox[t]{15cm}{\footnotesize 
What happens if $M_R$ has one or more zero eigenvalues? 
Obviously, in this case $M_R^{-1}$ does not exist, and the usual seesaw 
approximation fails. However, it can be readily modified to produce meaningful 
results. One can just go to the $\nu_R$ basis where $M_R$ is diagonal and 
include the lines and columns of $M_R$ that contain zero eigenvalues into a 
redefined matrix $m_L$. This situation is called the `singular seesaw', and it 
generally leads to the existence of pseudo-Dirac light neutrinos.
}}

\vspace*{1.5mm}

What can one say about the expected mass scale of the right-handed neutrinos 
$M_R$? Let us take 
the mass of the heaviest among the light neutrinos to be   
$m_\nu \sim 5\times 10^{-2}$ eV, as required by the data of the atmospheric 
and accelerator neutrino oscillation experiments under the assumption of the 
hierarchical neutrino masses. Then, assuming that the largest eigenvalue of 
the Dirac neutrino mass matrix is of the order of the electroweak scale, 
$m_D\sim 200$ GeV, and that $m_L\lesssim m_D^2/M_R$, 
from eq.~(\ref{eq:ss5}) we find $M_R\sim 10^{15}$ GeV. 
Interestingly, this is very close to the expected grand unification scale 
$m_{\rm GUT}\sim 10^{16}$ GeV. Thus, neutrino oscillations together with the 
seesaw mechanism of neutrino mass generation may be giving us an indication 
in favour of grand unification of weak, electromagnetic and strong 
interactions. 

The version of the seesaw mechanism discussed here, with heavy sterile 
neutrinos responsible for the small light neutrino masses, is sometimes 
called type I seesaw. There also exist other versions -- those with 
heavy $SU(2)_L$-triplet Higgs scalars (type II seesaw), heavy triplet 
fermions (type III seesaw), as well as other realizations of the seesaw 
mechanism, see ref.~\cite{numassReview} for a review. In all these cases 
neutrinos of definite mass are generically Majorana particles.

\subsection*{\label{sec:nuem}15.5\, Electromagnetic properties of Majorana 
neutrinos }

As neutrinos are electrically neutral, they have no direct coupling to 
the photon and their electromagnetic interactions arise entirely through loop 
effects (see ref.~\cite{GiuStud} for a review). 
It is interesting to compare the electromagnetic properties of Majorana 
neutrinos with those of Dirac neutrinos, which we discuss first.  
The matrix elements of the electromagnetic current $j^\mu(x)$ between the 
1-particle on-shell states of a Dirac neutrino (or any other Dirac fermion) 
can be written as 
\be
\langle \vec{p}',s'|j^\mu(x)|\vec{p},s\rangle=
e^{i q x}\,
\langle \vec{p}',s'|j^\mu(0)|\vec{p},s\rangle=
e^{i q x}\,
\bar{u}_{s'}(\vec{p}')\Lambda^\mu(q)u_s(\vec{p})\,,
\label{eq:em1}
\ee
where $q\equiv p'-p$, $u_s(\vec{p})$ and $u_{s'}(\vec{p}')$ are the 
free-particle plane wave spinors, and 
\be
\Lambda^\mu(q)=
F_Q(q^2)\gamma^\mu+F_M(q^2)i\sigma^{\mu\nu}q_\nu+F_E(q^2)\sigma^{\mu\nu}
\gamma_5 q_\nu+F_A(q^2)(q^2 g^{\mu\nu}-q^\mu q^\nu)\gamma_\nu\gamma_5\,.
\label{eq:em2}
\ee
Here $F_Q(q^2)$ and $F_M(q^2)$ are the electric charge and magnetic dipole 
form factors, while $F_E(q^2)$ and $F_A(q^2)$ are the electric dipole 
form factor and anapole form factor, respectively. 

\vspace*{1.5mm}
\colorbox{lightgray}
{\parbox[t]{15cm}{\footnotesize 
Unlike the magnetic and electric dipole moments, the anapole moment, first 
proposed in \cite{Zeld}, has no simple classical multipolar analogue. It can be 
modeled by a torus-shaped solenoid and therefore is sometimes called the 
toroidal moment. For a discussion of the properties of the anapole moment 
and its experimental manifestations see \cite{Dub,Esposito} and references 
therein. }}

\vspace*{1.5mm}

\noindent
The form of $\Lambda^\mu(q)$ in (\ref{eq:em2}) follows from the requirements 
of Lorentz covariance and electromagnetic current conservation.%
\footnote{There are alternative (but equivalent) forms of $\Lambda^\mu(q)$.
We prefer the one in (\ref{eq:em2}) because each term in it separately 
conserves the electromagnetic current. The same applies to eq.~(\ref{eq:em2a}) 
below.} 
For interactions with real photons the anapole form factor does not contribute. 
The hermiticity of the Hamiltonian of the electromagnetic interaction 
${\cal H}_{int}=j^\mu(x)A_\mu(x)$ implies that all the four form factors 
in eq.~(\ref{eq:em2}) are real.  
The vector-current form factors $F_Q(q^2)$ and $F_M(q^2)$ are parity 
conserving, while the axial-vector ones $F_E(q^2)$ and $F_A(q^2)$ 
violate parity. In addition, $F_E(q^2)$ violates CP invariance. 
As electroweak interactions that induce the effective neutrino electromagnetic 
current do violate parity (and possibly also CP), 
in general the form factors $F_A(q^2)$ and $F_E(q^2)$ need not vanish.

The charge form factor taken at zero squared 4-momentum transfer yields 
the electric charge of the Dirac particle, $F_Q(0)=Q$, whereas $F_M(0)$, 
$F_E(0)$ and $F_A(0)$ give, respectively, its anomalous magnetic moment 
$g-2$, electric dipole moment and anapole moment. 
The term $F_Q(q^2)\gamma^\mu$ in eq.~(\ref{eq:em2}) describes in the static 
limit ($q^2\to 0$) not only the electric charge interaction of the Dirac 
fermion, but also its normal magnetic moment. This can be seen from the 
Gordon identity 
\be
\bar{u}_{s'}(\vec{p}')\gamma^\mu u_s(\vec{p})=
\bar{u}_{s'}(\vec{p}')\left[\frac{p^\mu+p'^{\mu}}{2m}+\frac{i\sigma^{\mu\nu}
q_\nu}{2m}\right] u_s(\vec{p})\,.
\label{eq:Gordon}
\ee
The first term in the square brackets here corresponds to the convective 
part of the current, while the second one describes its spin part, i.e.\ the 
normal magnetic moment. The total magnetic moment of the particle 
is the sum of the normal and the anomalous ones. 

Because neutrinos 
have no electric charge (i.e.\ for them $F_Q(0)=0$), 
they do not have normal magnetic moments either.
Note that electric neutrality does not mean that the entire charge form factor 
$F_Q(q^2)$ vanishes. At small $q^2$ one can write
\be
F_Q(q^2)= F_Q(0)+F_Q'(0)q^2 +\dots\equiv 
F_Q(0)+\frac{1}{6}\langle r\rangle^2 q^2 +\dots\,.
\label{eq:chrad}
\ee
The quantity $\langle r^2\rangle$ characterizes the charge distribution 
within the particle and 
is called its charge radius.
It is in general different from zero even for neutral particles.   
Non-triviality of their charge distributions is related to the fact that 
interactions that `dress' the particles produce clouds of virtual 
particles of opposite charges and in general different configurations.  

Let us now discuss the electromagnetic properties of Majorana neutrinos 
\cite{SchVal2,PalWolf,Nieves,Kayser2,Shrock,KayGold,Kayser1}. 
For a Majorana particle all the electromagnetic form factors but one vanish 
identically, and the matrix element of the electromagnetic current takes the 
form 
\be
\langle \vec{p}'|j^\mu(0)|\vec{p}\rangle=\bar{u}(\vec{p}')[
F_A(q^2)(q^2 g^{\mu\nu}-q^\mu q^\nu)\gamma_\nu\gamma_5\,]u(\vec{p})\,.
\label{eq:em3}
\ee
That is, while the electromagnetic properties of a Dirac fermion are in 
general described by four form factors, for a Majorana particle only the 
anapole form factor survives. The simplest way to see this is to note that 
each term in eq.~(\ref{eq:em2}) can be viewed as emerging from the matrix 
element of the corresponding effective operator $\bar{\psi}\Gamma^\mu \psi$, 
where  $\psi$ is the free field operator and $\Gamma^\mu =(\gamma^\mu, 
\sigma^{\mu\nu}q_\nu, \sigma^{\mu\nu}\gamma_5q_\nu,\gamma^\mu\gamma_5)$, 
between the free one-particle states. For Majorana neutrinos 
from eqs.~(\ref{eq:C2}) and (\ref{eq:C2a}) we find 
\be
\bar{\psi}_k\gamma^\mu \psi_i=-\bar{\psi}_i\gamma^\mu \psi_k\,,
\qquad
~~~\bar{\psi}_k \gamma^\mu\gamma_5 \psi_i
=\bar{\psi}_i\gamma^\mu\gamma_5\psi_k\,,
\label{eq:cond1}
\ee
\be
\,~~~~\bar{\psi}_k\sigma^{\mu\nu}\psi_i=-\bar{\psi}_i\sigma^{\mu\nu} \psi_k\,,
\quad
\;\,~~\bar{\psi}_k \sigma^{\mu\nu}\gamma_5\psi_i=-\bar{\psi}_i\sigma^{\mu\nu}
\gamma_5\psi_k\,,
\label{eq:cond2}
\ee
where we have used the Majorana condition (\ref{eq:C4}). 
In the case when the fields $\psi_i$ and $\psi_k$ correspond to the same 
particle, i.e.\ $k=i$, from eqs.~(\ref{eq:cond1}) and (\ref{eq:cond2})
it follows that the only non-zero operator is the axial-vector one, 
$\bar{\psi}_i \gamma^\mu\gamma_5 \psi_i$, 
whereas all the other operators 
vanish identically. Electromagnetic current conservation then 
implies that the axial-vector operator can enter the matrix element 
(\ref{eq:em1}) only through the anapole interaction. 
This result has a simple interpretation. 
The charge radius and magnetic and electric dipole 
moments have opposite signs for neutrinos and antineutrinos and thus could be 
used to distinguish between them. 
Therefore, they  must vanish if neutrinos are Majorana fermions. At the same 
time, the anapole moment does not change its sign under the 
particle-antiparticle conjugation \cite{Zeld,Dub}, 
and so is allowed. Note that the vanishing of the charge, 
magnetic dipole and electric dipole form factors of Majorana neutrinos 
has a deep reason -- it is related to CPT invariance.

Expressions (\ref{eq:em1}) and (\ref{eq:em2}) describe the matrix elements 
of the electromagnetic current $j^\mu(x)$ between the 1-particle states 
of an individual neutrino. 
They are actually a special case of more general matrix elements of $j^\mu(x)$ 
between the states of neutrinos of different mass. 
These matrix elements have a form similar to eq.~(\ref{eq:em1}), 
except that the quantity $\Lambda^\mu(q)$ and the form factors are now 
matrices in the space of neutrino mass eigenstates. The requirements of 
Lorentz invariance and electromagnetic current conservation yield 
\be
\Lambda^\mu(q)_{ki}=
[F_1(q^2)_{ki}-\gamma_5 F_A(q^2)_{ki}](q^2 g^{\mu\nu}-q^\mu q^\nu)\gamma_\nu 
+[F_M(q^2)_{ki}-i\gamma_5 F_E(q^2)_{ki}]\,i\sigma^{\mu\nu}q_\nu\,.
\label{eq:em2a}
\ee
Eq.~(\ref{eq:em2}) corresponds to the diagonal elements of (\ref{eq:em2a}), 
with $F_Q(q^2)=F_Q(q^2)_i\equiv F_1(q^2)_{ii}q^2$ (note that for $i=k$ the 
expression $\bar{u}_{ks'}(\vec{p}')\gamma^\mu q_\mu u_{is}(\vec{p})$ vanishes 
identically since the spinors satisfy the Dirac equation). 
The off-diagonal matrix elements of the form factors in (\ref{eq:em2a}) are 
called the transition form factors. They describe transitions $\nu_i\to \nu_k$ 
caused by interaction of neutrinos with real or virtual photons or external 
electromagnetic fields. We will briefly discuss some of these processes  
towards the end of this section. 
Let us also note that, unlike for the diagonal elements of the form-factors,
hermiticity of the electromagnetic interaction Hamiltonian ${\cal H}_{int}$ 
does not by itself mean that the transition form factors are real. However, 
hermiticity of ${\cal H}_{int}$ combined with the assumption of CP 
invariance would 
require the form factors to be relatively real, i.e.\ for 
given $i,k$ all $F_a(q^2)_{ki}$ ({\footnotesize $a={1,A,M,E}$}) 
would be allowed to differ from their respective 
complex conjugates only by the same phase factor.

The expression for $\Lambda^\mu(q)_{ki}$ has the same form~(\ref{eq:em2a}) 
for Dirac and Majorana neutrinos, though in the Majorana case the form factors 
must satisfy some additional constraints. We have found that for Majorana 
neutrinos the diagonal matrix elements of the electromagnetic current contain 
only one non-zero form factor, $F_A(q^2)_{ii}$. This is an immediate 
consequence of eqs.~(\ref{eq:cond1}) and (\ref{eq:cond2}). However, for 
transition form factors the constraints are less severe. 
Eqs.~(\ref{eq:cond1}) and (\ref{eq:cond2}) then simply imply that the form 
factors $F_1(q^2)_{ki}$, $F_M(q^2)_{ki}$ and $F_E(q^2)_{ki}$ are antisymmetric 
with respect to the interchange of the indices $i$ and $k$, whereas 
$F_A(q^2)_{ki}$ is symmetric. Actually, this is a consequence of CPT 
invariance; one can readily check this by making use of the transformation 
properties of Majorana states under CPT given in Table~I and taking into 
account that the CPT transformation is anti-unitary and that the 
electromagnetic current $j^\mu(0)$ is CPT-odd.%
\footnote{In general, the symmetry or antisymmetry relations $F_a(q^2)_{ik}=\pm 
F_a(q^2)_{ki}$ should include an extra phase factor $\eta_{ki}$ related to the 
CPT-parities of the Majorana neutrinos $\nu_i$ and $\nu_k$. These CPT-parities, 
however, are not physically observable (unlike the CP-parities in the case when 
CP is conserved), and so one can set $\eta_{ki}=1$ without loss of generality.}

Another important point is that
non-vanishing transition electric dipole form factor $F_E(q^2)_{ki}$ would not 
necessarily signify leptonic CP violation. It is only in the case when both 
the transition magnetic dipole and 
electric dipole form factors $F_M(q^2)_{ki}$ and $F_E(q^2)_{ki}$ are 
simultaneously different from zero that one would have to conclude that CP 
is violated. In fact, 
if CP is conserved in the leptonic sector, 
the form factors 
$F_1(q^2)_{ki}$ and $F_M(q^2)_{ki}$ vanish when $\nu_i$ and $\nu_k$ have the 
same CP-parity (either $i$ or ${}-i$), 
whereas $F_A(q^2)_{ki}$ and $F_E(q^2)_{ki}$ vanish when $\nu_i$ and $\nu_k$ 
have opposite CP-parities \cite{KayGold}.   
This can be readily shown 
using the transformation properties of 
Majorana states under CP given in Table~I. 
If CP is violated, neutrinos do not possess definite CP-parities, and the 
simultaneous existence of non-zero $F_M(q^2)_{ki}$ and $F_E(q^2)_{ki}$ 
(or $F_1(q^2)_{ki}$ and $F_A(q^2)_{ki}$) with $k\ne i$ is allowed.

We have mentioned in sec.\ 15.1 that in the limit of vanishing neutrino 
mass Dirac and Majorana neutrinos become indistinguishable (as both actually 
become Weyl neutrinos). It is therefore interesting to see how the 
electromagnetic properties of Dirac and Majorana neutrinos converge in the 
massless limit. 
Let us first note that the vector and axial-vector operators 
$\bar{\psi}_i\gamma^\mu\psi_k$ and $\bar{\psi}_i\gamma^{\mu}\gamma_5\psi_k$ are 
chirality-preserving, while $\bar{\psi}_i\sigma^{\mu\nu}\psi_k$ and 
$\bar{\psi}_i\sigma^{\mu\nu}\gamma_5\psi_k$ are chirality-flipping: 
\be
\bar{\psi}_i\sigma^{\mu\nu}\psi_k=\bar{\psi}_{iL}\sigma^{\mu\nu}\psi_{kR}+
\bar{\psi}_{iR}\sigma^{\mu\nu}\psi_{kL}\,,
\label{eq:chir1}
\ee 
and similarly for $\bar{\psi}_i\sigma^{\mu\nu}\gamma_5\psi_k$. 
No left-right transitions can be induced by loop effects in the 
massless neutrino limit, and so the dipole form factors $F_M(q^2)$ and 
$F_E(q^2)$ must vanish in this limit identically. 

\vspace*{1.5mm}
\colorbox{lightgray}
{\parbox[t]{15cm}{\footnotesize 
This result is quite general and is easy to understand. If there is 
a loop diagram giving a contribution to the chirality-flipping neutrino 
magnetic or electric dipole form factors, then the same diagram with the 
external photon line removed will give a contribution to the neutrino 
mass term, 
which is also chirality-flipping. Thus, for massless neutrinos their magnetic 
and electric dipole form factors must vanish identically. 
(It is possible to devise symmetries that allow neutrino magnetic moments
but forbid neutrino masses \cite{Vol}, but such symmetries must be broken in 
the real world).  
}}

\vspace*{1.5mm}

\noindent
Thus, we are left with only vector and axial-vector operators. 
Next, we notice that for massless neutrinos 
$\bar{u}(\vec{p}')\gamma^\mu q_\mu u(\vec{p})=\bar{u}(\vec{p}')\gamma^\mu 
\gamma_5 q_\mu u(\vec{p})=0$. The quantity $\Lambda^\mu(q)_{ki}$ can therefore 
be written as 
\be
\Lambda^\mu(q)_{ki}=
F_1(q^2)_{ki}q^2\, \gamma^\mu+F_A(q^2)_{ki}q^2\, \gamma^\mu\gamma_5\,.
\label{eq:em2b}
\ee
The two terms here contain the neutrino charge form factor and the anapole 
form factor.  For Majorana neutrinos, the former is antisymmetric and the 
latter is symmetric with respect to the indices $i$ and $k$, while no such 
constraints  exist in the Dirac case. 
As massless neutrinos are chiral and 
$\gamma_5 u_{L,R}=\mp u_{L,R}$, the two terms in (\ref{eq:em2b}) are actually 
indistinguishable and merge into one, which is neither symmetric nor 
antisymmetric. Thus, the restrictions on the neutrino electromagnetic 
interactions that are specific to the Majorana case disappear in the limit 
$m_\nu\to 0$. 

The same conclusion can also be achieved in a different way 
\cite{Kayser2}. As follows from the Majorana condition 
(\ref{eq:C4}) (or equivalently from eq.~(\ref{eq:decompM})), 
for each loop diagram contributing to the electromagnetic vertex of a Dirac 
neutrino, in the Majorana case there is an additional diagram with all 
particles replaced by their $\hat{C}$-conjugates.%
\footnote{There will also be extra diagrams in the Majorana case if the sector 
of the model responsible for the Majorana mass generation contains charged 
particles. 
However, in the limit of vanishing neutrino mass which we consider now 
such diagrams can be we neglected. 
}
If the original diagram is 
caused by left-handed currents, then the $\hat{C}$-conjugate one is due to 
the right-handed interactions of the antiparticles (see eq.~(\ref{eq:C3})). 
The additional diagrams contribute to the electromagnetic vertices of massive 
Majorana neutrinos because Majorana fields contain both the left-handed 
and right-handed parts, the latter being $\hat{C}$-conjugates of the former. 
Decoupling of the left-handed and right-handed neutrino states in the massless 
limit means that the contribution of these additional diagrams to the 
amplitude of a given electromagnetic transition 
becomes negligible as $m_\nu\to 0$. 
Vanishing contributions of the diagrams that are specific 
to the Majorana case means that the electromagnetic properties of Majorana 
neutrinos converge to those of Dirac neutrinos 
when $m_\nu\to 0$.

In sec.\ 15.3 we pointed out that a pair of mass-degenerate Majorana neutrinos 
with maximal mixing and opposite CP-parities merges into a Dirac neutrino. The 
electromagnetic properties of such a pair should then be the same as those 
of a Dirac neutrino. This is indeed the case; in particular, the transition 
magnetic moment of such a Majorana pair becomes the usual magnetic moment of 
the Dirac neutrino.  We refer the reader to ref.~\cite{Nieves} for details.

{}From the above discussion it follows that the electromagnetic properties 
of massive Dirac and Majorana neutrinos are very different. Can 
this be used to find out whether neutrinos are Dirac or Majorana particles? 
To answer this question, we should first examine 
how the neutrino electromagnetic properties can manifest themselves.  
First, through the the photon exchange diagrams, they 
can contribute to the cross sections of $\nu f$ scattering, where 
$f$ is a charged lepton or a quark. In principle, such contributions can 
probe the neutrino magnetic and electric dipole moments, as well as 
the charge radius and anapole moment. Unfortunately, up to now experiment 
and observations   
(most notably, experiments on $\bar{\nu}_e e$ scattering with reactor 
antineutrinos as well as astrophysical data) have failed to discover  
neutrino electromagnetic properties and only produced upper limit on them 
\cite{GiuStud}. This is actually not surprising, as in the standard model 
and its simplest extensions 
the neutrino electromagnetic interactions are expected to be extremely weak. 
As an example, the standard-model prediction for the neutrino charge radius 
is $\langle r^2\rangle \sim 10^{-33}$ cm$^2$, whereas adding right-handed 
neutrinos to the model, for the diagonal magnetic moments of Dirac neutrinos 
one finds \cite{FuShr} 
\be
\mu_i\approx \frac{3 e G_F}{8\sqrt{2}\pi^2 }m_i\approx 3.2\times 
10^{-19}\Big(\frac{m_i}{\rm eV}\Big)\mu_B\,,
\label{eq:magnmom}
\ee 
where $e$ is the absolute value of the electron charge, $G_F$ is the Fermi 
constant, $m_i$ is the mass of the $i$th neutrino mass eigenstate and 
$\mu_B=e/2m_e$ is the electron Bohr magneton.
Similar expressions can be obtained for transition magnetic moments 
\cite{PalWolf,Shrock}. In addition, as we discussed above, 
the smallness of neutrino mass makes it very difficult to tell Majorana 
neutrinos from Dirac ones through their electromagnetic properties. In 
particular, it is difficult to distinguish experimentally the neutrino 
charge radius (which is non-zero only for Dirac neutrinos) from the anapole 
moment, which is the only non-vanishing diagonal electromagnetic moment of 
Majorana neutrinos.%
\footnote{It should also be noted that there are some difficulties in defining 
the neutrino charge radius in a gauge-invariant and process-independent way, 
see discussion in sec.\ 3.3 of~\cite{GiuStud}.}

As we discussed above, Dirac neutrinos can in general have both diagonal and 
transition dipole moments, whereas for Majorana 
neutrinos only transition dipole moments are allowed. For neutrinos of both 
types transition magnetic and electric dipole moments will cause radiative 
decays of heavier neutrinos into lighter ones, $\nu_i\to \nu_k + \gamma$. 
Although the rates of the radiative decay of Dirac and Majorana neutrinos are 
in general different, because of large uncertainties in the involved neutrino 
parameters it is impossible to 
establish the nature of neutrinos by measuring their radiative decay widths. 
However, the circular polarizations of the 
produced photons are very different in the Dirac and Majorana cases, and this 
is completely independent of the neutrino unknowns \cite{Shrock}. 
In addition, for polarized parent neutrinos, the angular distributions 
of the emitted photons are different for neutrinos of the two types 
\cite{LW,Shrock}. 
Thus, at least in principle, one could distinguish between the Dirac and  
Majorana neutrinos by measuring the 
polarization or angular distribution of the photons produced in radiative 
neutrino decay. 

Unfortunately, it is rather unlikely that neutrino radiative decays will ever 
be observed,  as they are doubly suppressed by the smallness of the neutrino 
magnetic moments (which implies small transition amplitude)   
and of neutrino mass (which means very small phase space volume of the decay). 
There is, however, some chance to observe radiative neutrino transitions if 
relatively heavy sterile neutrinos exist.  

Neutrino diagonal and transition dipole moments can in principle 
manifest themselves differently -- through 
neutrino spin precession in strong electromagnetic fields.
Such a process can be caused by the interaction of neutrino magnetic 
\cite{FuShr} or electric \cite{Okun} dipole moments with external fields. 
Transition dipole moments can give rise to  
spin-flavour precession, in which neutrino spin and flavour are flipped 
simultaneously 
\cite{SchVal2,VVO}. This process can be resonantly enhanced when neutrinos 
propagate in matter \cite{LM,A}. 

Let us compare spin and spin-flavour precessions of Dirac and Majorana 
neutrinos. It is convenient to introduce the matrix of neutrino 
electromagnetic moments \cite{VVO}
\be
\tilde{\mu}=\mu+i\epsilon\,,
\label{eq:tildemu}
\ee
where $\mu$ and $\epsilon$ are the hermitian matrices of neutrino magnetic 
dipole and electric dipole moments, respectively. 
In the flavour eigenstate basis the 
dipole moment couplings of neutrinos to an external electromagnetic field are 
described by the effective operators 
\be
\frac{1}{2}\big[\bar{\nu}_{\beta}\,\sigma^{\mu\nu}
(\mu-i\epsilon\gamma_5)_{\beta\alpha}
\nu_{\alpha}\big]F_{\mu\nu}+h.c. = 
\frac{\tilde{\mu}_{\beta\alpha}}{2}\,
\overline{\nu_{\beta R}}\,\sigma^{\mu\nu}
\nu_{\alpha L}
F_{\mu\nu}+h.c.
\quad\quad\,
\mbox{(Dirac)};\qquad
\label{eq:flip1}
\ee
\be
\frac{1}{2}\big[\bar{\nu}_{\beta}\,\sigma^{\mu\nu}
(\mu-i\epsilon\gamma_5)_{\beta\alpha}
\nu_{\alpha}\big]F_{\mu\nu}+h.c. = 
\frac{\tilde{\mu}_{\beta\alpha}}{2}\,\overline{(\nu_{\beta L})^c}\,
\sigma^{\mu\nu}\nu_{\alpha L}F_{\mu\nu}+h.c. 
\quad \mbox{(Majorana)},\,
\label{eq:flip2}
\ee
where $F_{\mu\nu}=\partial_\mu A_\nu-\partial_\nu A_\mu$ is the electromagnetic 
field tensor. 
Note that there is no extra factor 1/2 in eq.~(\ref{eq:flip2}) because in the 
Majorana case the matrix $\tilde{\mu}$ is antisymmetric.

It is instructive to look at the right-hand sides of eqs.~(\ref{eq:flip1}) 
and~(\ref{eq:flip2}), which reveal the nature of the involved neutrinos. 
Eq.~(\ref{eq:flip1}) means that in a transverse%
\footnote{Magnetic and electric dipole interactions of relativistic 
neutrinos with longitudinal (i.e.\ collinear with the neutrino momentum) 
magnetic fields are strongly suppressed \cite{FuShr,SchVal2}.}
external magnetic field e.g.\ a left-handed (active) electron neutrino 
$\nu_{eL}$ can be converted into a right-handed sterile neutrino of the same 
or different flavour. 
At the same time, in the Majorana case only flavour-off-diagonal 
transitions are allowed. For instance, for $\alpha=e$ and $\beta=\mu$ the 
interaction in eq.~(\ref{eq:flip2}) describes the transformation  
of an active left-handed electron neutrino $\nu_{eL}$ to the active 
right-handed muon neutrino state $\bar{\nu}_{\mu R}=(\nu_{\mu L})^c$ which is 
usually called the muon antineutrino. From this discussion it is clear that 
neutrino spin and/or spin-flavour precession lead to physically very different 
final states for Dirac and Majorana neutrinos and therefore could in principle 
be used to discriminate between them.  

Although the neutrino dipole moments are 
expected to be very small, the neutrino spin precession and spin-flavour 
precession can still occur with sizeable probabilities in extremely 
strong magnetic fields which may be present in astrophysical objects. In 
particular, in the Majorana neutrino case strong magnetic fields present 
in supernovae during the explosion stage may cause the resonantly 
enhanced conversion $\nu_e\to \bar{\nu}_\mu$. The resulting muon 
antineutrinos will then experience the usual flavour transitions on 
their way from the supernova to the Earth, converting them to electron 
antineutrinos. As a result, the overall neutrino transmutation chain 
$\nu_e\to \bar{\nu}_\mu\to \bar{\nu}_e$ will transform electron 
neutrinos into electron antineutrinos. Such a conversion of supernova 
$\nu_e$'s would have a very clear signature in the terrestrial detectors 
\cite{AF,AS}, 
provided that the supernova event occurs in our galaxy and that the 
transition magnetic moments of Majorana neutrinos $\mu\gtrsim 
10^{-14}\mu_B$.%
\footnote{Here we are assuming that at the resonance of spin-flavour 
conversion the supernova transverse magnetic field $B_{\perp r}$ 
can be as large as $\sim 10^{9}$ G \cite{AF}. The $\nu_e\to \bar{\nu}_e$ 
conversion efficiency depends on 
the product $\mu B_{\perp r}$.} 
While such relatively large values of $\mu$ are not easily achieved, 
they are predicted in some models and 
are not excluded by the current data 
and observations. At the same time, the $\nu_e\to \bar{\nu}_e$ 
conversion (which is a $\Delta L=2$ process) cannot occur if neutrinos 
are Dirac particles. Thus, future supernova neutrino experiments may 
shed some light onto the Dirac vs Majorana nature of neutrinos. Yet, the 
most practical means of disentangling these two neutrino types is 
probably neutrinoless double $\beta$-decay, which will be discussed in 
sec.\ 15.7.

\noindent
\subsection*{\label{sec:susy}15.6\, Majorana 
particles in SUSY}

In supersymmetric (SUSY) theories each boson has a supersymmetric 
partner which is a fermion and 
each fermion has a bosonic superpartner. Such theories predict the existence 
of a plentitude of Majorana fermions, which are supersymmetric partners of 
neutral bosons \cite{Nilles,HabKane,Drees}. These include the photino, 
as well as the gluino, zino and neutral higgsinos (the SUSY 
partners of the photon, gluon, $Z^0$-boson and of the neutral Higgs 
scalars, respectively). More precisely, since these particles can mix, what 
actually makes the Majorana fermions are the so-called neutralinos -- the 
linear superpositions of the above-mentioned particles that have definite 
masses.%
\footnote{We assume here that these Majorana particles are non-degenerate in 
mass. Otherwise two Majorana fermions can merge into a Dirac one, as discussed 
in sec.\ 15.3.3.}
In addition, if 
the spontaneous breaking of global supersymmetry occurs, there should exist 
the goldstino -- a massless neutral Goldstone fermion. In supergravity the 
goldstino is absorbed, through a supersymmetric analogue of the Higgs 
mechanism, into the gravitino, which is a massive spin-3/2 Majorana fermion 
(the SUSY partner of the graviton). In SUSY versions of the models where the 
so-called strong CP problem is solved through the existence of a light neutral 
pseudoscalar particle -- the axion --  
there is yet another Majorana fermion, the axino.

In SUSY models with conserved $R$-parity the lightest supersymmetric particle 
is stable. If it is neutral, it can be the so-called WIMP (weakly interacting 
massive particle) and play a role of the dark matter particle, i.e.\ account 
for the missing matter of the Universe \cite{Jung,Bertone,Feng2}. 
The lightest supersymmetric particle is then the lightest neutralino, the 
gravitino or the axino.%
\footnote{The role of a dark matter particle can also be played by a non-SUSY 
sterile Majorana neutrino, see \cite{Bertone,Feng2,Kus} and references 
therein.}   
Thus the dark matter problem, which is one of the most important problems of 
modern cosmology, may have its solution through the existence of a Majorana 
fermion.

\subsection*{\label{sec:exp}15.7\, Experimental searches for Majorana 
neutrinos and other Majorana particles}

\subsubsection*{15.7.1\, Neutrinoless 2$\beta$ decay and related processes}

\begin{figure}[h]
\begin{center} 
\begin{picture}(349,80)(-15,-40)
\ArrowLine(-10,25)(20,25)
\ArrowLine(-10,-85)(20,-85)
\Photon(20,25)(20, 0)3 4
\Text(15,12)[r]{\small $W$}
\Text(17,-110)[c]{$(a)$}
\ArrowLine(20,0)(50,10)
\ArrowLine(50,-10)(20,0)
\ArrowLine(50,-50)(20,-60)
\ArrowLine(20,-60)(50,-70)
\Photon(20,-60)(20,-85)3 4
\Text(15,-72)[r]{\small$W$}
\Text(55,10)[l]{\small$e_L$}
\Text(55,-73)[l]{\small$e_L$}
\Text(55,-49)[l]{\small$\nu_{eL}$}
\Text(55,-10)[l]{\small$\nu_{eL}$}
\Text(45,35)[r]{\small$u_L$}
\Text(50,-93)[r]{\small$u_L$}
\Text(-10,-93)[l]{\small$d_L$}
\Text(-10,35)[l]{\small$d_L$}
\ArrowLine(20,25)(50,25)
\ArrowLine(20,-85)(50,-85)
\ArrowLine(120,25)(150,25)
\ArrowLine(120,-85)(150,-85)
\Photon(150,25)(150, 0)3 4
\Text(155,12)[l]{\small$W$}
\Text(152,-110)[c]{$(b)$}
\ArrowLine(150,0)(180,0)
\ArrowLine(150,-30)(150,0)
\Text(150.5,-30)[c]{$\times$}
\ArrowLine(150,-30)(150,-60)
\ArrowLine(150,-60)(180,-60)
\Photon(150,-60)(150,-85)3 4
\Text(155,-72)[l]{\small$W$}
\Text(185,0)[l]{\small$e_L$}
\Text(185,-60)[l]{\small$e_L$}
\Text(155,-30)[l]{\small$m_L$}
\Text(145,-15)[r]{\small$\nu_{eL}$}
\Text(145,-45)[r]{\small$\nu_{eL}$}
\Text(175,35)[r]{\small$u_L$}
\Text(180,-93)[r]{\small$u_L$}
\Text(120,-93)[l]{\small$d_L$}
\Text(120,35)[l]{\small$d_L$}
\ArrowLine(150,25)(180,25)
\ArrowLine(150,-85)(180,-85)
\ArrowLine(250,25)(280,25)
\ArrowLine(250,-85)(280,-85)
\Photon(280,25)(280, 0)3 4
\Text(285,12)[l]{\small$W_L$}
\Text(282,-110)[c]{$(c)$}
\ArrowLine(280,0)(310,0) 
\ArrowLine(280,-25)(280,0)
\Text(280.9,-21)[c]{$\times$}
\ArrowLine(280,-40)(280,-25)
\Text(280.9,-42)[c]{$\times$}
\ArrowLine(280,-40)(280,-60)
\ArrowLine(280,-60)(310,-60)
\Photon(280,-60)(280,-85)3 4
\Text(285,-72)[l]{\small$W_R$}
\Text(315,0)[l]{\small$e_L$}
\Text(315,-60)[l]{\small$e_R$}
\Text(285,-21)[l]{\small$m_D$}
\Text(285,-42)[l]{\small$M_R$}
\Text(275,-10)[r]{\small$\nu_{eL}$}
\Text(275,-30)[r]{\small$\nu_{eR}$}
\Text(275,-50)[r]{\small$\nu_{eR}$}
\Text(305,35)[r]{\small$u_L$}
\Text(310,-93)[r]{\small$u_R$}
\Text(250,-93)[l]{\small$d_R$}
\Text(250,35)[l]{\small$d_L$}
\ArrowLine(280,25)(310,25)
\ArrowLine(280,-85)(310,-85)
\end{picture}
\end{center}
\vspace{2.2cm}
\caption{\small Some Feynman diagrams for the amplitudes of 
$2\beta$ decay. \label{fig:2beta}}
\end{figure}
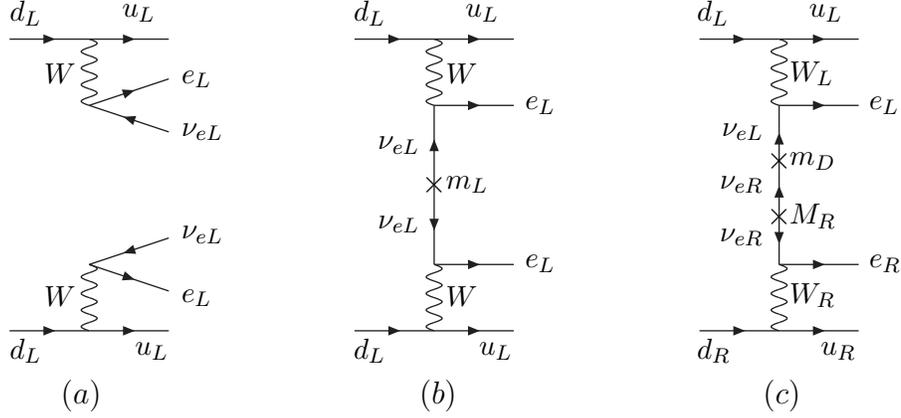

As was mentioned above, the most practical way of discriminating between Dirac 
and Majorana neutrinos seems to be by looking for neutrinoless double beta 
decay (see refs.~\cite{doi1,elvogel,Rode,Barabash1} for reviews). 
The usual double beta decay is the process in which a nucleus $A(Z,N)$ 
decays into an isobar with the electric charge differing by two units: 
\be
A(Z,N)\to A(Z\pm 2,N\mp 2)+2e^\mp +2\bar{\nu}_e(2\nu_e)\,.
\label{eq:2beta2nu} 
\ee
In such decays two neutrons of the nucleus are simultaneously converted into  
two protons, or vice versa. At the fundamental (quark) level, these are  
transitions of two $d$ quarks into two $u$ quarks or vice versa (see 
fig.~\ref{fig:2beta}a). Double beta decay is the process of the second order 
in weak interaction, and the corresponding decay rates are very low: typical 
lifetimes of the nuclei with respect to the $2\beta$ decay are $T\gtrsim   
10^{19}$ years. The processes (\ref{eq:2beta2nu}) are called 
$2\nu\beta\beta$ decays. Two-neutrino double beta decays with the emission of
two electrons ($2\beta^-$) were experimentally observed for a number of 
isotopes with the half-lives in the range $\sim 10^{19} - 10^{24}$ years 
\cite{Barabash1}; there are few candidate nuclei for 
$2\beta^+$ decay, and the experimental observation of this process is 
difficult because of the very small energy release ($Q$ values).

If neutrinos are Majorana particles, 
the lepton number is not conserved, and the neutrino emitted in one of the 
elementary beta decay processes forming the $2\beta$ decay can be absorbed in 
another (fig.~\ref{fig:2beta}b), leading to the neutrinoless double beta 
($0\nu\beta\beta$) decay \cite{Furry}:
\be
A(Z,N)\to A(Z\pm 2,N\mp 2)+2e^\mp \,.
\label{eq:2beta0nu} 
\ee
Such processes would have a very clear experimental 
signature: since the recoil energy of a daughter nucleus is negligibly small, 
the sum of the energies of the two electrons or positrons in the final state 
should be equal to the total energy release, i.e. should be represented by a 
discrete energy line. Therefore $0\nu\beta\beta$ decays could serve as a 
sensitive probe of the lepton number violation and Majorana nature of 
neutrinos. In some extensions of the standard model exotic modes of 
$0\nu\beta\beta$ decay are possible, e.g. decays with a Majoron emission
\cite{doi1,Rode}. In this case the sum of the energies of two electrons or 
positrons is not a discrete line, but 
the $2\beta$ energy spectra (as well as the single $\beta$-particle spectra)  
are expected to be different from those in the case of $2\nu\beta\beta$ decay.

Neutrinoless $2\beta$ decays break not only the lepton number; since the
absorbed $\nu_e$ or $\bar{\nu}_e$ has a `wrong' chirality, $0\nu\beta\beta$ 
decays also break chirality conservation. Therefore, 
if $0\nu\beta\beta$ decay is mediated by the standard weak interactions 
and exchange of light neutrinos,  
the amplitude of the process must be proportional to the neutrino mass. More 
precisely, as follows from fig.~\ref{fig:2beta}b, it is proportional to the 
$ee$-entry of the neutrino Majorana mass matrix, whose modulus is usually 
called $\langle m_{\beta\beta}\rangle$: 
\be
A(0\nu\beta\beta) \propto \Big|\sum_i U_{ei}^2 \, m_i\Big|\equiv 
\langle m_{\beta\beta}\rangle\,.
\label{eq:ampl}
\ee
Notice that this expression contains $U_{ei}^2$ rather than $|U_{ei}|^2$.  
If CP is conserved in the leptonic sector, the mixing matrix $U_{ai}$ 
can always be made real; however in this case the mass parameters $m_i$ 
in (\ref{eq:ampl}) (the eigenvalues of the neutrino mass matrix) can be of 
either sign, their relative signs being related to the relative CP parities 
of neutrinos. This means that in general significant cancellations between
various contributions to the sum in (\ref{eq:ampl}) are possible. As we 
discussed in sec.\ 15.3.3, a pair of Majorana neutrinos with equal physical
masses $|m_i|$, opposite CP parities and maximal mixing is equivalent to 
a Dirac neutrino. It is easy to see that such a pair does not contribute to 
the amplitude in (\ref{eq:ampl}) -- the contributions of the two components 
of the pair cancel exactly. Analogously, the contribution of a 
pseudo-Dirac neutrino to (\ref{eq:ampl}) would be strongly suppressed. 
Partial cancellations of contributions of different neutrino mass eigenstates 
to $\langle m_{\beta\beta}\rangle$ are also possible, of course, when CP is 
violated, i.e.\ when the leptonic mixing matrix is complex 
(with the convention that 
all neutrino masses are non-negative).  
In the case of just three usual light neutrino species 
eqs.~(\ref{eq:ampl}) and~(\ref{eq:U2}) yield 
\be
\langle m_{\beta\beta}\rangle=
\big|c_{13}^2 c_{12}^2 m_1\;+\;c_{13}^2 s_{12}^2 e^{2i\varphi_1} m_2
\;+\;s_{13}^2 e^{2i(\varphi_2-\delta_{CP})} m_3\big|.
\label{eq:mbetabeta}
\ee
By now, the neutrino oscillation experiments measured rather accurately the 
leptonic mixing angles $\theta_{12}$, $\theta_{23}$, $\theta_{13}$ and the 
neutrino mass squared differences $\Delta m_{21}^2$ and $|\Delta m_{31}^2|$. 
Global analyses of the data \cite{glob1,glob2,glob3} yield  
\be
\Delta m_{21}^2
\simeq 7.5\times 10^{-5}~\mbox{eV}^2\,,\qquad
|\Delta m_{31}^2|\simeq 2.4\times 10^{-3}~\mbox{eV}^2\,,\qquad 
\label{eq:glob1}
\ee
\be
\theta_{12}\simeq 33^\circ, \qquad \theta_{23}\simeq 40^\circ~~\mbox{or}~~
50^\circ, \qquad
\theta_{13}\simeq 9^\circ.
\label{eq:glob2}
\ee
At the same time, at present there is essentially no information on the 
CP-violating phases and the neutrino mass ordering (the sign of 
$\Delta m_{31}^2$), while for the absolute scale of the neutrino masses only 
upper limits exist: direct neutrino mass measurements in nuclear $\beta$-decay 
experiments and cosmology yield $m_i\lesssim {\cal O}(1)$ eV. With these data, 
it follows from eq.~(\ref{eq:mbetabeta}) that sizeable cancellations between 
the contributions of the different neutrino mass eigenstates to $\langle 
m_{\beta\beta}\rangle$ are possible only in the case of the so-called 
normal neutrino mass hierarchy, $m_1, m_2\ll m_3$. 

If, in addition to the usual three light neutrino species, there exist heavy 
neutrinos $N_i$, the 
active flavour-eigenstate neutrinos are linear superpositions of the 
left-handed components of both light and heavy neutrino mass eigenstates (see 
eq.~(\ref{eq:mix1})). Since the chirality-flipping part of the fermion 
propagator $m/(p^2-m^2)\simeq -1/m$ for $m^2\gg p^2$, 
the contribution of the diagram~\ref{fig:2beta}b with exchanges of heavy 
Majorana neutrinos to the amplitude of $0\nu\beta\beta$ decay is proportional 
to 
\be
\langle m_N^{-1}\rangle\equiv\Big|\sum_{i=4}^n
U_{ei}^2 m_i^{-1} \Big|.
\label{eq:mN}
\ee
Thus, one should distinguish effects of light and heavy Majorana neutrino 
exchanges. The latter requires the existence of extra neutrino species and can 
be considered as one of non-standard mechanisms of $0\nu\beta\beta$ decay. 
The effect of Majorana neutrino exchanges on  
lepton number violating processes is expected to be maximal when the mass of 
the exchanged neutrino is of the order of the characteristic energy of the 
process. This applies not only to $0\nu\beta\beta$ decay but to all $\Delta 
L=2$ processes, including those considered in sec.\ 15.7.2 below.

\begin{figure}[htb]
\begin{center}
{\includegraphics[width=8.2cm,height=2.7cm]{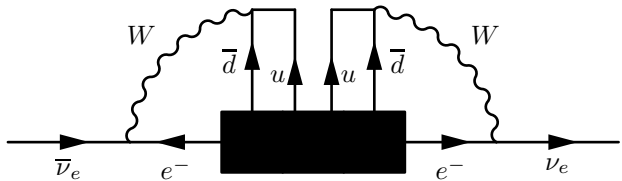}}
\end{center}
\caption{The black box argument for Majorana neutrino mass \cite{SchVal3}.}
\label{fig:bb}
\end{figure}

In extensions of the standard model, such as the left-right symmetric, SUSY or 
grand unification models, additional mechanisms of $0\nu\beta\beta$ decay are 
possible, in which 
the process is mediated by right-handed currents, SUSY particles or leptoquarks 
(see, e.g., \cite{Rode}). 
One of the diagrams contributing to the amplitude of $0\nu\beta\beta$ decay 
in left-right symmetric theories is shown in fig.~\ref{fig:2beta}c. It may 
appear that no Majorana mass $m_L$ of $\nu_L$ is necessary in such models, 
i.e.\ $0\nu\beta\beta$ decay can occur even if $m_L=0$ and the neutrinos are 
Dirac particles, or even if they are massless. This is, however, incorrect: in 
all models in which $0\nu\beta\beta$ decay occurs, the Majorana masses of 
$\nu_L$ must be different from zero. 
An elegant `black box' proof of this statement was presented in \cite{SchVal3}.
In fig.~\ref{fig:bb} the black box represents an unspecified mechanism by which 
two $d$-quarks can be converted to two $u$-quarks and two electrons, with no 
accompanying neutrinos. 
Next, we make use of the crossing symmetry to transform the initial-state 
$d$-quarks to the final-state $\bar{d}$-quarks, join the $u\bar{d}$ lines to 
produce $W$-bosons and then attach the other ends of the $W$-boson lines to 
the electron lines to produce neutrinos. This yields the diagram corresponding 
to the effective operator $\bar{\nu}_{eL} (\nu_{eL})^c$ describing the 
$\bar{\nu}_{eR}\to \nu_{eL}$ transition, i.e.\ the Majorana mass term for 
$\nu_e$.   
Thus, no matter what mechanism causes $0\nu\beta\beta$ decay, observation of 
this process would constitute an unambiguous proof that neutrinos are Majorana 
particles.

If neutrinos are of Majorana nature, then in addition to the usual 
$0\nu\beta\beta$ decay~(\ref{eq:2beta2nu}), some related processes should 
occur, such as e.g.\  $e_B+A(Z,N)\to A(Z-2,N+2)+e^+$ (neutrinoless electron 
capture) or $2e_B+A(Z,N)\to A(Z-2,N+2)^* \to A(Z-2,N+2)+X$ (neutrinoless 
double electron capture) \cite{Winter}. Here $e_B$ stands for a bound atomic 
electron and $A(Z-2,N+2)^*$ denotes the excited state of the $A(Z-2,N+2)$ 
atom with two holes in the atomic $1S$ orbit, which 
then de-excites with the emission of atomic $X$-rays, Auger electrons, 
etc..\ These processes are always energetically allowed if the usual $2\beta^+$ 
decay of $A(Z,N)$ is allowed, and in some cases may also be allowed 
even if $A(Z,N)$ is stable. The neutrinoless double electron capture 
may be resonantly enhanced provided that the 
total energy release is very close to the 
excitation energy of the $A(Z-2,N+2)^*$ atomic 
state. The search for isotopes with suitable atomic mass differences and 
excitation energies of the daughter atoms is currently under 
way \cite{Eliseev}. 

Other processes related to $0\nu\beta\beta$ decay have been discussed, such as 
muon conversion on nuclei, $\mu^-+A(Z,N)\to A(Z-2,N+2)+e^+$ or $\mu^-+A(Z,N)
\to A(Z-2,N+2)+\mu^+$. Just like the $0\nu\beta\beta$ decay, such conversion 
processes break the total lepton number $L=L_e+L_\mu+L_\tau$ by two units 
(note, however, that the $(\mu^-,e^+)$ conversion conserves $L_e-L_\mu$). 
If these processes are mediated by exchanges of light or heavy Majorana 
neutrinos,  their expected rates are too small  to render them observable in 
a foreseeable future \cite{domin,missimer}. Thus, the muon conversion 
processes may be of interest only if they are dominated by non-standard 
mechanisms.

It is customary to discuss the available data of $0\nu\beta\beta$ 
decay experiments as well as expected sensitivities of the ongoing and future 
experiments in 
terms of the limits on the half-lives of the parent nuclei $T_{1/2}^{0\nu}$ and 
interpret them in terms of the effective mass parameter $\langle 
m_{\beta\beta}\rangle$ defined in~(\ref{eq:ampl}). It should, however, be 
remembered that such an interpretation makes sense only when the standard 
diagram of fig.~\ref{fig:2beta}a with exchange of light Majorana neutrinos 
is the sole (or the dominant) contribution to the amplitude of the process, 
in which case $T_{1/2}^{0\nu}\propto 1/\langle m_{\beta\beta}\rangle^2$. 
Otherwise, there is no or little connection between these two quantities, and 
the masses of light neutrinos cannot be directly probed in experiments on 
$0\nu\beta\beta$ decay.%
\footnote{
Similar argument applies to the contributions of heavy Majorana neutrino 
exchanges to $0\nu\beta\beta$ decay rates, which depend on the quantity 
$\langle m_N^{-1}\rangle$ defined in eq.~(\ref{eq:mN}), and the possibility to 
probe the masses of heavy neutrinos. 
}
If $0\nu\beta\beta$ decay is dominated by non-standard mechanisms, 
experiments can only give upper limit on the parameter $\langle m_{\beta\beta}
\rangle$ and therefore on the Majorana masses of the neutrino mass 
eigenstates.  
However, as was stressed above, even if the Majorana neutrino mass gives 
negligible contribution to $0\nu\beta\beta$ decay, an observation of this 
process would be an unambiguous proof of the Majorana nature of neutrinos. 
Experimentally, one can in principle distinguish between different 
mechanisms of $0\nu\beta\beta$ decay by studying the properties 
of the decay products, e.g.\ angular correlations of the two produced 
$\beta$-particles, or by looking for processes related to $0\nu\beta\beta$ 
decay (see e.g.\ discussion in sec.\ 6 of ref.~\cite{Rode}).

Neutrinoless double beta decay has been actively  
searched for but up to now it has not been experimentally discovered.%
\footnote{There is one positive claim of observation of $0\nu\beta\beta$ 
decay of $^{76}$Ge by part of the Heidelberg-Moscow Collaboration 
\cite{KK}. However, this result has been subject to criticism (see e.g.\ 
\cite{Barabash1} and references therein) 
and is now strongly disfavoured (at 99\% C.L.) by non-observation of 
$0\nu\beta\beta$ decay by GERDA \cite{gerda}.}
The available data allow to put upper bounds on the effective 
Majorana neutrino mass $\langle m_{\beta\beta}\rangle$. The best current limits 
come from the EXO-200 experiment on $0\nu\beta\beta$ decay of $^{136}$Xe 
\cite{exo} and GERDA experiment with $^{76}$Ge \cite{gerda} (90\% C.L.):
\be
\langle m_{\beta\beta}\rangle < 0.14 - 0.38 ~\mbox{eV}~~~(\mbox{EXO-200})\,;
\qquad
\langle m_{\beta\beta}\rangle < 0.2 - 0.4 ~\mbox{eV}~~~(\mbox{GERDA})\,,
\label{eq:upper}
\ee
where the ranges are due to the uncertainties in the values of the nuclear 
matrix elements. 
The most promising current and forthcoming experiments 
are expected to be sensitive to the values 
$\langle m_{\beta\beta}\rangle \gtrsim 0.03$ -- $0.1$ eV \cite{Barabash1}; as 
follows from (\ref{eq:glob1}) and 
(\ref{eq:glob2}), they will be able to explore the Majorana neutrino mass 
only if neutrinos are quasi-degenerate in mass 
($m_1\sim m_2\sim m_3$) or have 
the inverted mass hierarchy ($m_3\ll m_1, m_2$). 

\vspace*{1.5mm}
\colorbox{lightgray}
{\parbox[t]{15cm}{\footnotesize
How does it work? As an example, assume that from 
independent experiments  
(e.g.\ neutrino oscillations plus cosmology) 
we know that neutrino masses obey the inverted hierarchy, i.e.\ $m_3\ll m_1, 
m_2$. From eq.~(\ref{eq:mbetabeta}) it then follows that 
the quantity $\langle m_{\beta\beta}\rangle$ cannot be smaller than  
$\langle m_{\beta\beta}\rangle_{\rm min}\approx c_{13}^2 \cos 2\theta_{12}
\sqrt{|\Delta m_{31}^2|}(\sim 0.02$ eV). If from a $0\nu\beta\beta$ 
experiment an upper limit on $\langle m_{\beta\beta}\rangle$ is inferred which 
is smaller than $\langle m_{\beta\beta}\rangle_{\rm min}$, this would rule 
out Majorana nature of neutrinos, barring destructive interference with 
some non-standard $0\nu\beta\beta$ decay mechanisms. 
}}
\vspace*{1.5mm}

\noindent
If the neutrino masses obey the 
normal mass hierarchy, probing the Majorana neutrino mass through 
$0\nu\beta\beta$ decay will be problematic and will in any case require 
multiton-scale detectors and very low backgrounds. 
It would be extremely difficult to uncover Majorana vs 
Dirac nature of neutrinos 
in this case, unless an efficient non-standard mechanism of $0\nu\beta\beta$ 
decay is at play.

\subsubsection*{15.7.2\, Other lepton number violating processes}

The Majorana nature of neutrinos can be revealed not only through neutrinoless 
double $\beta$-decay, but also through other lepton number violating processes. 
One of such processes -- neutrino spin-flavour precession in strong external 
magnetic fields -- was discussed in sec.\ 15.5. Here we shall briefly 
discuss two of the other types of $\Delta L=2$ processes, rare particle decays 
and like-sign dilepton production at accelerators. 

If neutrinos are Majorana particles, they should mediate rare $\Delta L=2$ 
particle decays, such as $K^+\to\pi^-e^+e^+$, $K^+\to\pi^-\mu^+\mu^+$, 
and similar decays of charged $B$ and $D$ mesons. 
The typical Feynman diagrams for such processes are essentially the same as 
the one in fig.~\ref{fig:2beta}b, except that some other quarks may be 
involved. 
However, the number of the parent particles in the case of rare meson decays 
is suppressed by a huge factor of order of the Avogadro number $N_A$ as 
compared to those in $2\beta$-decay experiments. 
Therefore, rare meson decays cannot compete with neutrinoless $2\beta$-decay 
in unraveling the neutrino nature when the exchanged Majorana 
neutrinos are very light or very heavy. Still, rare decays can provide 
tighter limits on the Majorana neutrino masses in the region of the order of 
the energy release of the corresponding process  
(for rare kaon decays, a few hundred MeV).

\begin{figure}[htb]
\begin{center}
{\includegraphics[height=3.7cm]{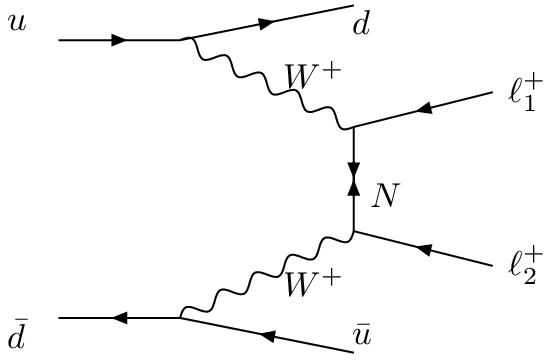}}
\hspace*{1.8cm}
{\includegraphics[height=3.7cm]{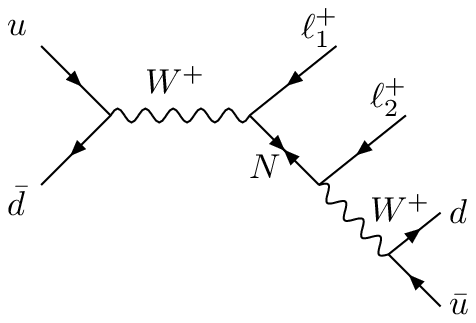}}
\end{center}
\caption{
Representative 
Majorana neutrino exchange diagrams contributing to like-sign dilepton 
production. Left: $W$-boson fusion with $t$-channel $N$ exchange. Right: 
$s$-channel $W$-exchange diagram with production and subsequent decay of $N$.}
\label{fig:diagr}
\end{figure}

In accelerator-based experiments the Majorana nature of neutrinos can be tested 
in the processes with like-sign dilepton production as well as in related 
reactions.
The basic $\Delta L=2$ processes are in this case $W^\pm W^\pm\leftrightarrow 
\ell_1^\pm \ell_2^\pm$, where $\ell_{1,2}=e, \mu$ or $\tau$. A typical reaction 
that can be studied at hadron colliders 
(first discussed in \cite{KS} in the context of right-handed currents) is 
$pp\to \ell_1^\pm \ell_2^\pm X$. Majorana neutrinos contribute to the amplitude 
of this process through the diagrams of fig.~\ref{fig:diagr} and similar 
diagrams with other quarks in the final state. 

\noindent
Here the left diagram is similar to diagram \ref{fig:2beta}b that gives the 
standard contribution to the amplitude of $0\nu\beta\beta$ decay, whereas 
the right diagram corresponds to production and subsequent decay of a 
virtual or real Majorana neutrino $N$. If real Majorana neutrinos are 
kinematically accessible, the dilepton production process can be resonantly 
enhanced. 
 
Lepton number violating rare decays and like-sign dilepton production 
processes have been actively looked for experimentally at accelerators, but 
no signals have been found so far. This allowed one to put important 
constraints on the properties of Majorana neutrinos. For a detailed and 
comprehensive discussion of these (and other) $\Delta L=2$ processes 
as well as of the effects of Majorana neutrinos on electroweak precision 
observables we refer the reader to ref.~\cite{Atre} 
(see also \cite{Rode} and references therein).

There have also been extensive searches for the SUSY Majorana particles both at 
accelerators and in dark matter detectors. For recent discussions of these 
experiments see e.g.\ \cite{Feng1,Kaz,Apr}. Unfortunately, up to now no 
unambiguous evidence for such particles has been obtained.

\subsection*{\label{sec:leptog}15.8\, Baryogenesis through leptogenesis and 
Majorana neutrinos}

In addition to providing a simple and natural way of explaining the observed 
smallness of the neutrino mass, the seesaw mechanism of the neutrino mass 
generation brings with it a free bonus: It furnishes a very simple and 
attractive mechanism of producing the observed baryon asymmetry of the 
Universe (BAU). The term baryon asymmetry simply reflects the fact that the 
observed Universe is made predominantly of matter rather than of equal amount 
of matter and antimatter (there are very stringent constraints on the 
antimatter abundance in the Universe \cite{dolgov}). The ratio of the net 
baryon number to photon number in the Universe is now measured very accurately 
\cite{Planck}:
\be
\eta \equiv \frac{N_B-N_{\bar{B}}}{N_\gamma}=(6.04\pm 0.08)
\times 10^{-10}\,,
\label{eq:baryon}
\ee
where $N_B$, $N_{\bar{B}}$ and $N_\gamma$ are the number densities of baryons,
antibaryons and photons 
at the present epoch. The observed 
BAU could not have resulted from an initial state of the Universe with $B\ne 
0$, as any such pre-existing asymmetry would have been diluted to an absolutely 
negligible level during the stage of the accelerated expansion of the Universe 
predicted by the cosmic inflation \cite{inflation} 
(which is the standard paradigm now). Thus, the BAU should have been generated 
dynamically in the post-inflationary epoch. 
 
Under what conditions such a dynamical generation of the BAU can occur? 
These conditions were actually formulated by Sakharov in 1967 \cite{Sakharov}:
(i) baryon number violation; (ii) C- and CP-violation; 
(iii) deviation from thermal equilibrium. 
The first two conditions are necessary for the baryon asymmetry to be produced 
in the first place; the last condition 
ensures that the BAU 
produced in some processes  
is not destroyed by the inverse processes. 

\vspace*{1.5mm}
\colorbox{lightgray}
{\parbox[t]{15cm}{\footnotesize 
As an illustration, consider a process $X\to Y+b$, where $X$ denotes an initial 
state with zero baryon number, $Y$ stands for a set of final-state 
particles with vanishing net baryon number 
and $b$ represents the produced excess baryons. Then, 
if condition (i) is not met, the process  $X\to Y+b$ just does not take place.
If either C or CP is conserved, the processes $X\to Y+b$ and $\bar{X}\to 
\bar{Y}+\bar{b}$ occur at the same rate, and no net baryon number is produced 
(provided that the initial state of the system contained equal numbers 
of $X$ and $\bar{X}$
or that $X=\bar{X}$). 
If the system is in thermal equilibrium, the processes $X\to Y+b$ and 
$Y+b\to X$ occur at the same rate (which is also true, of course, for 
$\bar{X}\to \bar{Y}+\bar{b}$ and $\bar{Y}+\bar{b}\to \bar{X}$), and the baryon 
asymmetry produced in direct processes is washed out by the inverse ones.
}}
\vspace*{1.5mm}

\noindent
All these conditions 
are actually satisfied in the standard model of particle physics, though 
the amount of CP violation in this model is insufficient and, most 
importantly, the deviation from thermal equilibrium is far too small to 
account for the measured value of the BAU (\ref{eq:baryon}) \cite{RS}. 
The Sakharov's conditions are 
fulfilled and the successful generation of the BAU is possible in many 
extensions of the standard model, such as grand unification theories and SUSY 
models \cite{RS,Cline}. Here we concentrate on the so-called 
baryogenesis via leptogenesis \cite{FY}, which is built-in in the seesaw 
mechanism of the neutrino mass generation and does not require any new physics 
besides the existence of heavy Majorana neutrinos. We just outline the 
mechanism here; for details and ramifications we refer the reader to a  
comprehensive review \cite{DNN}.

To produce the phenomenologically acceptable mass spectrum of the usual light 
neutrinos, the seesaw mechanism should include at least two heavy 
electroweak-singlet (i.e.\ sterile) Majorana neutrinos $N_i$. The same number 
of $N_i$'s turns out to be sufficient for the generation of the BAU. 
The mechanism works as follows. First, a lepton number $L_0$ is produced 
through the out-of-equilibrium $L$- and $(B-L)$-violating decays of the 
$N_i$'s. The produced lepton number is then reprocessed into the baryon number 
by the so-called sphaleron processes (hence the name baryogenesis through 
leptogenesis). 

Let us consider this in more detail. The singlet neutrinos 
$N_i$ are actually not completely sterile: they cannot have gauge interactions 
in the standard model, but can have the usual Yukawa couplings 
$h_{\alpha i}^*\bar{\ell}_\alpha N_{iR}H+h.c.$%
\footnote{Following the tradition, for the right-handed components of the 
singlet neutrinos $N_i$ we use here the notation $N_{iR}$ rather than 
$\nu_{iR}$ that was used in secs.\ 15.3 and 15.4.} 
with the lepton doublets 
$\ell_\alpha=(\nu_{\alpha L}, e_{\alpha L})^T$ and the Higgs field 
$H=(H^0, H^-)^T$, 
which are allowed by the electroweak gauge symmetry.  
The Yukawa couplings result in decays of $N_i$ into the usual leptons and 
the Higgs particles. Since the $N_i$'s are Majorana particles, their decay 
proceeds in a lepton number violating way, i.e.\ they can decay both via 
$N_i\to \ell_\alpha \bar{H}$ 
and through the CP-conjugate channel $N_i\to \bar{\ell}_\alpha{H}$. 
If the Yukawa couplings $h_{\alpha i}$ are complex, CP is not conserved in the 
leptonic sector, and the rates of the above decay modes are in general 
different: $\Gamma(N_i\to \ell_\alpha \bar{H})\ne \Gamma(N_i\to 
\bar{\ell}_\alpha H)$. This leads to the production of a non-zero net lepton 
number. Note that CP-violation manifests itself through the 
interference between the tree-level and 1-loop Feynman diagrams describing the 
$N_i$ decay; at the tree level the decay rates are proportional to 
$|h_{\alpha i}|^2$, and the complexity of the Yukawa constants does not reveal 
itself. The parameter that describes the generation of the lepton asymmetry 
in the decay of $N_i$ is 
\be
\epsilon_{i}=\sum_\alpha \frac{\Gamma(N_i\to \ell_\alpha \bar{H})-
\Gamma(N_i\to \bar{\ell}_\alpha H)}{\Gamma(N_i\to \ell_\alpha \bar{H})
+\Gamma(N_i\to \bar{\ell}_\alpha H)}=\frac{1}{8\pi(h^\dag h)_{ii}} 
\sum_{j\ne i}{\rm Im}
[(h^\dag h)_{ij}^2]\,g(x_j)\,,
\label{eq:epsil1}
\ee
where $x_j\equiv M_j^2/M_i^2$ and $g(x)$ is model dependent. In the standard 
model 
\be
g(x)=\sqrt{x}\left[\frac{2-x}{1-x}-(1+x)\ln\left(\frac{1+x}{x}\right)\right].
\label{eq:gx}
\ee
In~(\ref{eq:epsil1}) we for simplicity summed over the flavours of 
final-state leptons (note that flavour effects may actually be important, see 
the discussion in sec.\ 9 of \cite{DNN}). Eq.~(\ref{eq:epsil1}) is 
valid only when the mass differences of the heavy singlet Majorana neutrinos 
are large compared with their decay widths, $|M_j-M_i|\gg \Gamma_i+\Gamma_j$; 
the opposite case, which leads to resonant leptogenesis \cite{PilUnd}, 
requires a special consideration.

The deviation from thermal equilibrium is provided by the expansion of the 
Universe, the rate of which is given by the Hubble parameter
\be
H(T)=1.66\sqrt{g_*}\,T^2/M_{\rm Pl}\,.
\label{eq:H}
\ee
Here $T$ is the temperature of the Universe, $M_{\rm Pl}=1.2\times 10^{19}$ 
GeV is the Planck mass and $g_*$ is the number of relativistic degrees of 
freedom in the thermal bath (in the standard model $g_*=106.75$). 
If the processes that create and destroy some particles are fast compared 
with the Hubble expansion rate $H(T)$, they equilibrate particle 
distributions, otherwise the thermal equilibrium is not achieved. For the 
singlet neutrino $N_i$ the condition of deviation from thermal equilibrium 
requires that at the time of the $N_i$ decay ($T\sim M_i$) the 
decay rate $\Gamma_i=\frac{(h^\dag h)_{ii}}{8\pi}M_i$ be smaller than the 
Hubble rate: $\frac{(h^\dag h)_{ii}}{8\pi}M_i\lesssim 1.66\sqrt{g_*}\,
\frac{T^2}{M_{\rm Pl}}|_{T\sim M_i}$. Thus, the lightest singlet neutrino 
$N_i$ is typically the last one to go out of equilibrium in the course of 
the expansion and cooling of the Universe. Therefore the lepton asymmetry 
produced in decays of heavier Majorana neutrinos is washed out by the 
processes involving $N_1$, and the net lepton asymmetry of the Universe is 
produced in the decays of $N_1$.%
\footnote{Under certain circumstances decays of heavier singlet neutrinos 
can also be important, see sec.\ 10.2 of \cite{DNN} and references therein.}
The out-of-equilibrium condition for these decays can be rewritten as 
\be
\tilde{m}_1\equiv \frac{(h^\dag h)_{11} v^2}{M_1} \lesssim 8\pi\cdot 
1.66\sqrt{g_*}\,\frac{v^2}{M_{\rm Pl}}
\simeq 1.1\times 10^{-3}~{\rm eV}\,, 
\label{eq:mstar}
\ee
where $v=174$ GeV is the Higgs VEV. 
In the opposite case the produced lepton asymmetry is strongly washed out. 
It is interesting to note that, since the Dirac-type neutrino mass matrix 
$m_D=h^T v$, the left-hand side of (\ref{eq:mstar}) is 
roughly of the same order of magnitude as the masses of the light active 
neutrinos predicted by the seesaw mechanism,%
\footnote{
If the matrix $h$ were real, for hierarchical masses of $N_i$
the left-hand side of (\ref{eq:mstar}) would have been approximately 
equal to the trace of the mass matrix of light neutrinos, i.e.\ to the sum of 
the light neutrino mass eigenvalues.}
whereas the right-hand side 
is close to the light neutrino mass scale that follows from 
the oscillation experiments assuming that neutrino masses are hierarchical. 
Typically, one expects 
the left-hand side of (\ref{eq:mstar}) to slightly exceed its 
right-hand side, leading to a moderate washout of the lepton asymmetry.

How can the produced 
lepton number be converted into the baryon one? 
In the standard model the baryon and lepton numbers are conserved at the tree 
level but are violated at 1-loop level by the so-called chiral anomalies.
The anomalies of the baryon and lepton number currents are the same, and 
so $B-L$ is exactly conserved, 
but the sum $B+L$ is not. Although the $(B+L)$-violating processes are 
strongly suppressed at zero temperature, the situation at high temperatures 
is different \cite{KRS}. The standard model predicts the existence of 
topologically non-trivial configurations of the gauge and Higgs fields (called 
sphalerons) which violate $B+L$ with the rate $\Gamma_{\rm sph}$ that exceeds 
the Hubble rate for $100~{\rm GeV} \lesssim T\lesssim 10^{12}~{\rm GeV}$. 
In this temperature interval the sphaleron processes can efficiently wash out 
$B+L$ and thus reprocess the lepton asymmetry into the baryon one. In a 
somewhat simplified way, the reprocessing mechanism can be described as 
follows. 

Assume that at a time $t=0$ a net lepton number $L_0$ is produced, while the 
initial baryon number $B_0=0$. Noting that $B$ and $L$ can be represented as 
linear combinations of $B-L$ and $B+L$ and that $B+L$ is exponentially 
suppressed with time by the sphaleron processes, for the values of $L$ and $B$ 
at a time $t\ge 0$ we have 
\be
L(t)=-\frac{1}{2}(B-L)_0+\frac{1}{2}(B+L)_0 e^{-\Gamma_{\rm sph}t}\,,
\nonumber
\ee
\be
B(t)=\frac{1}{2}(B-L)_0+\frac{1}{2}(B+L)_0 e^{-\Gamma_{\rm sph}t}\,. 
\label{eq:LB}
\ee
Thus, at the times $t\gg \Gamma_{\rm sph}^{-1}$ we have $L\simeq L_0/2$, 
$B\simeq -L_0/2$, i.e.\ we end up with non-zero baryon number. 
A realistic calculation, which takes into account that  
only left-handed quarks and leptons are coupled to the $W$-boson field, 
yields (in the standard model) $B=-(29/78) L_0$ rather than $-L_0/2$. In 
addition, one should carefully take into account the processes that wash out 
the lepton asymmetry, such as inverse $N_1$ decays and $2\to 2$ scattering 
processes. This is usually done by solving a system of Boltzmann equations or 
quantum kinetic equations.  
As a result, one finds that for hierarchical masses of heavy singlet neutrinos 
the observed value of the BAU can be generated provided that the mass 
of the lightest among the heavy Majorana neutrinos $N_1$ satisfies $M_1\gtrsim 
10^8$ GeV. For quasi-degenerate in mass heavy neutrinos the viable baryon 
asymmetry can be achieved, through the resonant leptogenesis, even 
for $M_i$ as small as $\sim 1$ TeV \cite{PilUnd}.

In the discussed baryogenesis mechanism all three Sakharov's conditions are 
satisfied. The baryon number violation is provided by the combination of 
$L$ (and $B-L$) violation in decays of heavy Majorana neutrinos and $B+L$ 
violation by the sphaleron processes. C-violation follows from the chiral 
nature of the Yukawa couplings, while CP-violation is a consequence of the 
complexity of the corresponding coupling constants. Finally, the condition of 
deviation from thermal equilibrium 
is met because in certain ranges of the parameters   
the rates of decay and inverse decay of the heavy Majorana neutrinos 
(as well as the rates of other $L$-violating processes) do not exceed 
significantly the Hubble expansion rate.

What we discussed above was baryogenesis via leptogenesis in type I seesaw.
Similar mechanisms work in the case of type II and type III seesaw \cite{DNN}.
There also exists an alternative leptogenesis mechanism \cite{ARS,Shap}, 
in which the lepton asymmetry is generated in CP-violating oscillations of 
the heavy Majorana neutrinos $N_i$ rather than in their decays. The produced 
asymmetry is then communicated from the $N_i$'s to the usual leptons through 
their Yukawa couplings, and the reprocessing of the lepton number to the 
baryon number proceeds through the sphaleron processes in the usual way.  
 
In all the discussed versions of baryogenesis through leptogenesis the 
Majorana nature of the singlet neutrinos plays a crucial role. There also 
exist leptogenesis scenarios with Dirac neutrinos (see sec.\ 10.4 of 
\cite{DNN} and references therein), but they are based on more complicated 
and less economical models.

\subsection*{\label{sec:misc}15.9\, Miscellaneous}
Here we collect 
a few assorted remarks on Majorana particles. 

It is usually said 
that in e.g.\ nuclear $\beta^-$ decay an electron antineutrino 
$\bar{\nu}_e$ is emitted, while positron production in $\beta^+$ decay is 
accompanied by the emission of an electron neutrino $\nu_e$, 
and we know that these are distinct particles. Does that mean that we have 
already established that neutrinos are Dirac particles and 
$\nu_e\ne (\nu_e)^c$? Not really. 
The point is that the charged-current weak interactions are chiral, so that 
only left-handed particles and their right-handed $\hat{C}$-conjugates can be 
emitted or absorbed. In the Dirac case, this means that only the left-handed 
component of the Dirac field $\nu_e=\nu_{eL}+\nu_{eR}$ (as well as 
the right-handed component $(\nu_{eL})^c\equiv \bar{\nu}_e$ of 
$(\nu_e)^c=(\nu_{eL})^c+(\nu_{eR})^c$) take part in the interactions, while 
$\nu_e$ and $(\nu_e)^c$ are indeed different particles. In the Majorana case 
we have $\nu_e=\nu_{eL}+(\nu_{eL})^c$, and both chiral components of the field 
participate in weak interactions.  What we call $\nu_e$ and $\bar{\nu}_e$ are 
in this case merely the left-handed and right-handed components of the same  
Majorana field of the electron neutrino. They are to a very good accuracy 
distinct because neutrinos we deal with are always highly relativistic, and 
the transitions between their left-handed and right-handed components are 
suppressed by the factor $(m_\nu/E)^2\ll 1$. Thus, the role of the lepton 
number, which is conserved in the Dirac case, is played 
for relativistic Majorana neutrinos 
by chirality, which is nearly conserved. This illustrates once again the 
point we have already made more than once  
-- the smallness of the neutrino mass 
makes it very difficult to discriminate between Dirac and Majorana neutrinos.

Can one still tell these two neutrino types apart by studying neutrino 
propagation under extreme conditions, such as e.g.\ very high densities and/or 
strong magnetic fields which are expected to be present in stellar 
environments? This question was studied in \cite{Esp}, and the answer 
unfortunately turns out to be essentially negative.

It is a well known but not yet completely understood fact that electric 
charge is quantized. Possible explanations include the existence 
of the magnetic monopole and grand unification of particles and forces. 
It turns out, however, that electric charge quantization can be understood 
even outside these frameworks if neutrinos are Majorana particles \cite{BM}.
In the minimal standard model with no right-handed (singlet) neutrinos $\nu_R$, 
charge quantization is a consequence of the hypercharge assignment of the 
particles that follows from the requirement of the anomaly cancellation. The 
cancellation of anomalies, in turn, is necessary for internal consistency of 
the theory. However, the minimal 
standard model is not realistic in the sense that neutrinos are massless in it. 
It therefore in any case has to be amended by a neutrino mass generating 
mechanism. If one adds right-handed singlet neutrinos to the standard model, 
there are essentially two possibilities. First, one imposes a lepton number 
conservation which allows only Dirac mass terms for neutrinos. In this case 
the anomaly cancellation condition no longer leads to electric charge 
quantization. If no lepton number conservation is imposed, massive neutrinos 
turn out to be Majorana particles. In this case anomaly cancellation 
always results in electric charge quantization \cite{BM}. The authors of 
\cite{BM} have also studied a wide class of non-grand-unified extensions of 
the standard model which allow massive neutrinos, and found that in virtually 
all cases the Majorana nature of neutrinos led to electric charge quantization, 
whereas for Dirac neutrinos no such quantization occurred. Thus, the observed 
quantization of electric charge in Nature may have its explanation through 
the existence of Majorana neutrinos.

It is conceivable that our (3+1)-dimensional world is actually embedded in 
a space-time of higher dimensionality; in particular, higher-dimensional 
space-times appear in Kaluza-Klein, supergravity and superstring models. 
{}From the point of view of applications to condensed-matter physics, it may 
also be interesting to consider space-times of lower dimensionality. 
Can Majorana particles live in such unconventional space-times? The answer is 
yes, but not in all of them. For $d$-dimensional space-times with $d-1$ 
space-like and one time-like dimensions, massive Majorana fermions can exist 
only if $d=2,3$ and 4 {\em mod} 8 (see e.g.\ sec.\ 2 of \cite{Pil} and 
references therein). Massless self-conjugate fermions can live in the 
space-times of the same dimensionality, and in addition in $d=8$ and 9 
{\em mod} 8 dimensions.%
\footnote{Note that massless spin-1/2 fermions admit more freedom in the 
definition of the the particle-antiparticle conjugation operation: the matrix 
${\cal C}$ that enters eq.~(\ref{eq:C}) may be defined either through the 
usual relation ${\cal C}^{-1}\gamma^\mu {\cal C}=-\gamma^{\mu T}$ or through 
${\cal C}^{-1}\gamma^\mu {\cal C}=+\gamma^{\mu T}$.}
These results can also be extended to the case of $n>1$ time-like dimensions.

\subsection*{\label{sec:concl}15.10\, 
Summary and conclusions }

The possibility of existence of fermions which are their own antiparticles is 
certainly the most famous and arguably the most important result obtained by 
Ettore Majorana. Extensions of the standard model typically predict neutrinos 
to be massive Majorana particles. There are some experimental hints in favour 
of possible existence of extra neutrino species (on top of the already known 
$\nu_e$, $\nu_\mu$ and $\nu_\tau$); if exist, they are very likely Majorana 
particles. The Majorana nature of neutrinos would imply lepton number 
violation -- a very interesting phenomenon which is now being intensely 
searched for experimentally. 

Possible existence of heavy electroweak-singlet Majorana neutrinos provides 
us, through the seesaw mechanism, with a natural and elegant explanation of 
the smallness of the masses of the usual neutrinos. Heavy (or relatively 
heavy) Majorana neutrinos furnish very simple and attractive mechanisms for 
generating the observed baryon asymmetry of the Universe. Majorana particles 
are abundant in SUSY models. 
Majorana fermions can play a role of the dark matter particles and thus 
provide a solution of one of the most important problems of modern cosmology.
Majorana neutrinos may hold a clue to the understanding of electric charge 
quantization observed in Nature. 

If Majorana particles exist, they should have special properties with 
respect to C-, CP- and CPT-transformations and  
possess very peculiar electromagnetic properties. By studying them we 
may be able to learn a great deal about the fundamental properties of particles 
and their interactions.  

Particle-like excitations of Majorana nature have been 
found in condensed-matter systems (see chapter 14 of this book). 
However, very active direct and indirect searches for Majorana neutrinos and 
other fundamental Majorana particles 
in many laboratories in the world have up to now brought no fruit. 
This should not discourage us too much -- 
just remember that it took us over 40 years to discover neutrino oscillations 
after their possibility had first been proposed! After all,
the idea of Majorana fermions is so elegant and attractive that Nature just 
could not have missed the opportunity to create them.


\end{document}